\documentclass[12pt]{article}
\usepackage{amsmath,amssymb,amsthm,amsfonts,setspace,fullpage,graphicx,xcolor,mathabx,paralist,nicefrac,comment,hyperref,booktabs,multirow,float,caption}
\usepackage[multiple]{footmisc}
\usepackage{thmtools}
\usepackage{thm-restate}

\makeatletter
\@ifundefined{newcounteralias}{}{%
  \renewcommand\thmt@autorefsetup{%
    \expandafter\def\csname\thmt@envname autorefname\expandafter\endcsname
      \expandafter{\thmt@thmname}}%
}
\makeatother
\usepackage{bbm}
\usepackage{apptools}
\usepackage{geometry}
\usepackage{sectsty}
\usepackage{tikz}
\usetikzlibrary{arrows.meta,positioning,calc}
\usepackage{adjustbox}
\usepackage{enumitem}
\setlist[description]{style=nextline}

\usepackage[style=authoryear,citestyle=authoryear-comp,natbib,maxcitenames=3,maxnames=5,uniquename=false,doi=false,giveninits=true]{biblatex}
\subsectionfont{\fontsize{13}{16}\selectfont}
\renewbibmacro{in:}{}
\DeclareFieldFormat{pages}{#1}
\renewbibmacro*{journal+issuetitle}{%
 \usebibmacro{journal}%
 \setunit*{\addcomma\addspace}%
 \iffieldundef{series}
  {}
  {\newunit
   \printfield{series}%
   \setunit{\addspace}}%
 \usebibmacro{volume+number+eid}%
 \setunit{\addspace}
 \usebibmacro{issue+date}%
 \setunit{\addcolon\space}%
 \usebibmacro{issue}%
 \newunit}
\AtEveryBibitem{
  \clearfield{issue}
  \clearfield{number}
  \clearfield{issn}
}
\uspunctuation
\hypersetup{colorlinks=true,linkbordercolor=red,linkcolor=blue,citecolor=blue}
\DeclareMathOperator*{\argmax}{argmax}

\newcommand{\pE}[2]{\mathbb{E}_{#1}\left[#2\right]}

\newcommand{\KL}{\mathrm{KL}}
\newcommand{\kl}{\mathrm{kl}}
\newcommand{\TV}{\mathrm{TV}}
\newcommand{\Dir}{\mathrm{Dir}}
\newcommand{\Bad}{\mathrm{Bad}}
\newcommand{\1}{\mathbbm{1}}
\newcommand{\norm}[1]{\left\lVert #1\right\rVert_1}
\DeclareMathOperator{\Opt}{Opt}
\usepackage{needspace,fix-cm}

\newtheorem{theorem}{Theorem}

\newtheorem{corollary}{Corollary}

\newtheorem{example}{Example}

\newtheorem{lemma}{Lemma}

\newtheorem{proposition}{Proposition}

\theoremstyle{definition}

\newtheorem{assumption}{Assumption}
\newtheorem{remark}{Remark}

\makeatletter
\renewenvironment{proof}[1][\proofname]{ \par\pushQED{\qed}\normalfont \topsep6\p@\@plus6\p@\relax
 \trivlist\item[\hskip\labelsep\bfseries#1\@addpunct{.}] \ignorespaces
}{ \popQED\endtrivlist\@endpefalse
}
\makeatother

\makeatletter
\newcommand{\pushright}[1]{\ifmeasuring@#1\else\omit\hfill$\displaystyle#1$\fi\ignorespaces}
\newcommand{\pushleft}[1]{\ifmeasuring@#1\else\omit$\displaystyle#1$\hfill\fi\ignorespaces}
\makeatother

\AtAppendix{
 \counterwithin{lemma}{section}
 
}
\AtAppendix{
 \setstretch{1.2}
}
\AtAppendix{
 \setlength{\abovedisplayskip}{10pt}
 \setlength{\belowdisplayskip}{10pt}
 \setlength{\abovedisplayshortskip}{8pt}
 \setlength{\belowdisplayshortskip}{8pt}
}

\makeatletter
\newrobustcmd*{\parentexttrack}[1]{
 \begingroup
 \blx@blxinit
 \blx@setsfcodes
 \blx@bibopenparen#1\blx@bibcloseparen
 \endgroup}
\makeatother
\AtEveryCite{  }

\title{Patient Learning in Endogenous Environments\thanks{We thank Daniel Clark for extremely helpful suggestions and NSF grant 2417162 for financial support. The proofs in this paper were generated using GPT-6.2 Astra and Claude Fable 5.2. In a series of interactive conversations, we  prompted these models  to generalize the initial result and  look for counterexamples.  We also consulted Refine, which suggested an improved proof  of our main result. The LLM's  also helped us   connect the results to the economic theory literature. We read and edited all of the text several times, and accept responsibility for any remaining mistakes.
}}
\author{Drew Fudenberg \and Kevin He}
\date{November 2026}
\begin{document}
\hypersetup{pageanchor=false}
\maketitle
\begin{abstract}
\singlespacing
In learning problems that generalize multi-armed bandits, we give conditions under which the regret of Bayesian-optimal policies vanishes uniformly over the true parameter as the agent becomes patient. Uniformity matters when the parameter being learned itself depends on patience, as in our three applications. In steady-state learning in games, play approaches Nash equilibrium as agents become patient and long-lived, at any relative rates. This extends earlier results that required lifetimes to grow faster than patience. As buyers become patient, the optimal profit of a seller whose product characteristics buyers cannot observe approaches what she could earn against informed buyers, and optimal product designs maximize social surplus in the limit. In congestion networks, stationary learning flows approach Wardrop equilibrium as travelers become patient and long-lived. For affine mean delays, we also bound welfare losses and show that fixed marginal-cost tolls yield efficiency in the limit.
\end{abstract}
\thispagestyle{empty}
\newpage \setcounter{page}{1}
\hypersetup{pageanchor=true}

\onehalfspacing

\section{Introduction}

A more patient agent has more reason to experiment, so we expect her payoff to approach the full-information benchmark as her patience increases. In many economic applications, however,  the parameter she learns about is determined by others' behavior, which itself depends on patience. For example, suppose a seller chooses product characteristics that buyers learn about through repeated purchases. The seller’s profit-maximizing choice depends on buyers’ purchasing behavior, which in turn depends on their patience and hence their willingness to experiment with unfamiliar products. Similarly, suppose commuters learn about expected travel times on different routes from their own experiences. These travel times depend on commuters’ aggregate route choices, which in turn depend on their patience: more patient commuters may be more willing to risk delays today to learn about alternative routes for future commutes.  In each case increasing the learners' patience  changes  the underlying parameter they seek to learn, so convergence to the full-information payoff for each fixed parameter does not imply convergence as patience increases.

We therefore study uniform convergence of payoffs over a given parameter space as the agent grows patient. If the worst-case gap to the full-information payoff over the parameter space uniformly vanishes with patience, then payoffs converge along any sequence of true parameters, including ones that depend on patience. We do this in a setting that generalizes multi-armed bandits.  Unlike a standard bandit, Section~\ref{sec:common} posits that the agent is learning about a set of ``sources'' and different actions may generate observations from overlapping subsets of sources; the parameter is the collection of source distributions. For example, an agent learning how a population of opponents play an extensive-form game is uncertain about the distributions of opponent actions at each information set; these are the sources. The agent learns by being matched with a random opponent, choosing a contingent plan in the game tree, and observing the realized path of play.  Different plans can reach the same opponent information set, so observations under one plan help evaluate another. Similarly, a trip through a network reveals the travel time on each traversed edge. Routes that share  edges provide information about one another. 

Section~\ref{sec:uniform} shows that in this environment, Bayesian-optimal payoffs approach the full-information benchmark uniformly over the true parameter as patience grows. In a standard finite-outcome bandit,  convergence in expectation over the prior follows from the Bayesian optimality of the agent's behavior,   as in \citet{aghion1991optimal}. However, this expectation can converge even if payoffs remain low at parameters with little prior weight, so it does not imply convergence at each parameter in the support, let alone uniform convergence. We obtain our uniform convergence  under a broad class of priors that can include correlated beliefs about different sources, and  convergence holds even when true distributions  lie on the boundary of the prior support.   The same convergence holds when the horizon used to evaluate
welfare is at least a fixed positive fraction of the agent's
optimization horizon. This covers buyers learning about
product characteristics chosen by a seller who may be more
or less patient than they are (Section~\ref{sec:design}). 

Our principal application of the uniform convergence result is to steady-state  models of learning in games 
(Section~\ref{sec:steady}).    Agents form populations that correspond to player roles in an extensive-form stage game. In each period, agents are  matched with
new opponents to  play the stage game,  learn about opponents' play from their own match outcomes, 
 and are replaced
by newborn agents when they exit. In a setting with deterministic lifetimes, \citet{Fudenberg1993} show that as agents grow more patient and long-lived -- but become long-lived suitably more quickly than they become patient -- steady-state behavior converges to a Nash equilibrium. In a related setting with geometric lifetimes, \citet{fudenberg2018type} show the same conclusion when patience and expected lifetimes both grow, but in an iterated limit where expected lifetimes grow more quickly. The question about whether agents also converge to Nash equilibrium if their patience grows more quickly than their lifetimes has remained open. Theorem~\ref{thm:nash} and Proposition~\ref{prop:det-nash} establish that learning outcomes  converge to Nash equilibrium in this joint limit with \emph{any} relative growth rate of patience and lifetime, in the geometric and deterministic lifetime models respectively. Mirroring the general uniform convergence results, Nash convergence holds for a broad class of prior beliefs that allow subjective correlations about opponents' behavior at different information sets.

Two examples in Section~\ref{sec:fail} show why our results cannot be  generalized to arbitrary full-support prior beliefs. In an ordinary bandit, there is a prior with an atom under which every optimal policy chooses an inferior arm forever at some fixed true parameter. In a three-player game, we construct priors that have continuous, strictly positive interior densities and independence across opponents but fail our requirements on the degree of prior correlation about behavior at different information sets for the same opponent. We find that these priors support  a non-Nash steady-state outcome at every patience and lifetime, because dependence across information sets of the same opponent makes observed behavior increasingly discourage experimentation elsewhere.  

 The uniform bound also applies to hidden product design (Section~\ref{sec:design}) and
congestion (Section~\ref{sec:congestion}). A seller chooses product characteristics and a constant price. A long-lived Bayesian buyer observes the price but can only learn about the characteristics by repeatedly purchasing and experiencing the product. As the buyer becomes patient, the seller's optimal profit approaches the informed monopoly benchmark where the buyer also observes the product characteristics, and asymptotically optimal products approach the surplus-maximizing set. The seller may be more or less patient than the buyer,
provided her discount horizon remains at least a fixed
positive fraction of the buyer's.  In a Bernoulli example with a common discount factor, optimal
quality approaches zero as the agents become myopic and $1/2$
as they become patient, with the technology and prior fixed.

In congestion networks, travelers move through a network and are initially uncertain about the delay distributions on different roads.  Their collective route choices determine
these delays. The uniform bound implies that stationary learning flows approach Wardrop equilibrium as travelers become patient and long-lived. For the standard benchmark where average delay on an edge is an affine function of its traffic load, total travel time is no more than $4/3$ the social optimum, plus a learning-error term that vanishes with patience. With optimal marginal-cost tolls on the roads, total travel time converges to the social optimum with patience. 

The proof in Appendix~\ref{sec:proof} first bounds regret averaged
over the posterior by comparing optimal behavior with a policy
that experiments and then chooses the action with the highest
sample mean payoff. Our prior restriction ensures that, with
uniformly high probability, every posterior retains positive mass
near the truth. Comparing true and posterior predictive
probabilities over short epochs then bounds regret at each true
parameter, uniformly over evaluation discount factors.

The classical discounted bandit is a special case of our model where
each arm reveals a draw from its own source. With independent
arms, the Gittins index gives an optimal policy
\citep{Gittins1979}, but at a fixed discount factor even an
optimal policy can settle on an inferior arm with positive
probability \citep{rothschild1974pricing}. For Thompson sampling, \citet{RussoVanRoy2014,BubeckLiu2013}
bound prior-averaged regret, while \citet{AgrawalGoyal2013}
gives frequentist finite-time regret bounds. For Gaussian arms,
\citet{Lattimore2016} gives finite-time regret bounds for the
finite-horizon Gittins strategy and extends them to the
Bayesian-optimal policy in the two-arm case. Our
Corollary~\ref{cor:bandit} shows that in finite-outcome bandits
satisfying Assumption~\ref{asst:1}, discounted regret vanishes
uniformly as patience grows, including at boundary parameters
and for every optimal policy. Theorem~\ref{thm:learning}
allows actions to share sources and beliefs about those sources
to be correlated. It also makes the convergence uniform over
bounded payoff assignments on the same trees and allows the
evaluation horizon to differ from the optimization horizon,
provided it is at least a fixed fraction as long.

Our observation trees are  related to combinatorial
semi-bandits with probabilistically triggered arms, where an
action reveals outcomes from several sources and these
outcomes can determine which additional sources are observed
\citep{ChenWangYuanWang2016,WangChen2017}. These papers derive
finite-time regret bounds for upper-confidence-bound
algorithms under restrictions on expected rewards, whereas
we study Bayesian-optimal policies under a fixed prior.

Our results also relate to uniform posterior concentration. \citet{diaconis1990uniform} establishes posterior concentration
around empirical frequencies for $\phi$-positive priors:
every ball of radius $\varepsilon$ centered in the simplex
has prior probability at least $\phi(\varepsilon)>0$.
The concentration bound depends on $\phi$ and holds uniformly
over samples and priors satisfying this condition.
\citet{fudenberg2023pathwise} requires the same lower bound, but 
only for balls centered in the prior's support,  and establishes
concentration on distributions that approximately minimize
Kullback--Leibler divergence from the empirical distribution.
Our bounded-density comparison with a product of Dirichlet
distributions implies such a prior-mass bound, but also
restricts dependence across sources, which lets us show that  a patient  optimizing agent experiments enough that their  payoff converges uniformly to its full-information level.  

\section{The Learning Model }\label{sec:common}
There are finitely many actions $s\in S$ and  sources $h=1,\ldots,M$, with finite outcome sets $A_h$ and unknown parameters $\theta_h\in\Delta(A_h)$ associated with each source $h$. We omit singleton sources and assume $M\ge1$. Conditional on $\theta$, all source draws are independent, within and across periods. Each action is associated with a known finite  tree. Each non-terminal node of the tree is either a source whose outgoing edges correspond to the source's outcomes, or a chance node with known randomizations over its  outgoing edges.  Each source appears at most once on any path, and a tree may use no source at all. Each terminal node of the tree has a known payoff in $[-U,U]$ for some $U>0$. When the agent chooses an action, she traverses down the associated tree, drawing an outcome from $A_h$ according to the distribution $\theta_h$ whenever she encounters a source $h$ and drawing an outgoing edge  from the known randomization whenever she encounters a chance node. The agent observes the entire path and receives the payoff at the terminal node. 

The agent starts with a prior belief $q_0$ over $\theta = (\theta_1, ..., \theta_M)$ and chooses an action in each period, updating her belief from  observations. To state our assumptions on the prior, we start from independent Dirichlet priors across sources, multiply  their joint density over $\theta$ by another function $f(\theta)$  bounded above and bounded away from zero, then re-normalize into a density. Since $f(\theta)$ need not be the product of marginals, this allows the prior beliefs about different sources to be correlated.

\begin{assumption}\label{asst:1} The prior $q_0$ has a measurable
density $f$ relative to a product of Dirichlet distributions:
\begin{equation}\label{eq:prior}
 \begin{gathered}
 q_0(d\theta)=f(\theta)D(d\theta),\qquad D=\bigotimes_h\Dir(\alpha_h),\\
 0<c_-\le f\le c_+<\infty,\qquad \alpha_{ha}>0.
 \end{gathered}
\end{equation}
\end{assumption}
\begin{remark}Independent Dirichlet priors satisfy Assumption~\ref{asst:1} with $f=1$. Any correlated prior density $q_0$ that is bounded above and bounded away from zero automatically satisfies  Assumption~\ref{asst:1}, as we can take $\alpha_{ha} = 1$ for every $h, a$ so that the reference Dirichlet prior $D$ is the uniform distribution.  Assumption~\ref{asst:1} further allows for correlated prior densities that vanish to zero along the boundary of the parameter space, provided they do so at a rate comparable to the product of Dirichlet distributions. For example, let two binary sources have success probabilities $0\le p_1,p_2 \le 1$ and  consider the joint prior density  $q_0(p_1, p_2)=1+\rho(2p_1-1)(2p_2-1)$ for some  $0<|\rho|<1$. This prior satisfies Assumption~\ref{asst:1} with respect to the uniform distribution on $[0,1]^2$ and we can take $f=q_0$, noting that $f$ lies between $c_- = 1-|\rho|$ and $c_+ = 1+|\rho|$. 
\hfill $\diamond$\end{remark}

Assumption~\ref{asst:1}
includes the priors with densities
bounded above and away from zero used in the proofs of
\citet{Fudenberg1993}. It also nests the prior class of \citet{fudenberg2017rare} which assumes priors are independent across sources, and that each marginal  satisfies 
``Condition~$\mathcal P$'': the ratio of the source's prior density
to some Dirichlet density is uniformly continuous and bounded
away from zero on the interior of the simplex.  
Assumption~\ref{asst:1} drops the continuity requirement
and permits dependence across sources, while retaining
upper and lower bounds on the joint density relative
to a product of Dirichlet distributions.
Condition~$\mathcal P$ for each marginal  does not
impose this joint restriction; Example~\ref{ex:1} shows
why it matters.
Section~\ref{sec:fail} compares our prior restriction with others 
in related work.

\Needspace{7\baselineskip}
\paragraph{Payoffs and regret.} The agent starts with prior $q_0$, chooses an action in each period, and maximizes the $\beta$-discounted expected payoffs for $0 \le \beta < 1$. A policy maps finite histories of past actions and observations into a possibly random action this period, and we write $\Opt_\beta(q_0)$ for the set of all optimal policies for prior $q_0$ and discount factor $\beta$.  

We measure learning by expected discounted payoff loss relative
to full information. The discount factor used to evaluate this
loss may differ from the one governing the agent's choices.
Let $u(s,\theta)$ be the expected payoff of action $s$ under parameter $\theta$, let $u^*(\theta)=\max_s u(s,\theta)$, and write $\ell(s,\theta)=u^*(\theta)-u(s,\theta)$ for the loss of taking action $s$ under $\theta$.  For a policy $\sigma$, let $0 \le \lambda < 1$
be the discount factor used to evaluate payoffs, and define the (normalized) discounted regret of the policy under parameter $\theta$ by
\begin{equation}\label{eq:regret}
 R_{\lambda}(\theta,\sigma)
 =(1-\lambda)\mathbb{E}_\theta^\sigma\!\left[\sum_{t\ge0}\lambda^t\ell(s_t,\theta)\right].
\end{equation}

We call $1/(1-\beta)$ the \emph{optimization horizon} and
$1/(1-\lambda)$ the \emph{evaluation horizon.} The former governs
the agent's choices; the latter governs how we average payoff
losses when measuring regret. Our results allow the evaluation horizon to be shorter or
longer than the optimization horizon, provided the ratio
of the evaluation horizon to the optimization horizon
remains bounded away from zero.

\section{Uniformly Vanishing Regret}\label{sec:uniform}
This section states our general result. The  associated proof is in Appendix~\ref{sec:proof}.

Write $\mathcal G_U$ for the set of all assignments of payoffs  in $[-U,U]$ to the terminal nodes of the action trees,
$u_g(s,\theta)$ for the resulting expected payoff, and
$\Opt_\beta(q_0;g)$ for the optimal policies under $g$.
Write $R^g_{\lambda}$ for regret under payoff assignment $g$.

\begin{theorem}\label{thm:learning}
For every fixed learning problem and prior satisfying \eqref{eq:prior} and every fixed $C_0\ge1$,
\begin{equation}\label{eq:uniform}
 \lim_{\beta\to1}\
\sup_{\substack{\lambda\in[0,1)\\
                 1-\lambda\le C_0(1-\beta)}}\
 \sup_{g\in\mathcal G_U}\ \sup_\theta
 \ \sup_{\sigma\in\Opt_\beta(q_0;g)}
 R^g_{\lambda}(\theta,\sigma)=0.
\end{equation}
The supremum includes parameters on the boundary of the parameter space.
\end{theorem}
Theorem~\ref{thm:learning}  shows that patient Bayesian optimality
yields vanishing per-period regret uniformly over the true parameter. It allows for the worst-case selection of any evaluation horizon at least $1/C_0$ times the optimization horizon, any terminal-node payoff assignments, and any optimization tie-breaking rule for every patience level $\beta$ (in addition to the worst-case true parameter $\theta$),  which is  useful in applications where these aspects of the environment endogenously change with the agent's patience.

The proof bounds the frequency of actions whose payoff loss at
the true parameter exceeds a fixed $\bar\ell>0$. We first bound
dynamic regret under the agent's beliefs. A policy that samples
each action many times and then chooses the action with the highest sample
mean has a small normalized expected discounted loss relative to full
information when the agent is patient, so an optimal policy has no
greater loss. 

The difficulty is that this bound averages over the posterior,
whereas Theorem~\ref{thm:learning} concerns each fixed true
parameter. Our prior restriction ensures that, with uniformly
high probability, every posterior assigns at least a fixed
positive mass $\kappa$ to a neighborhood of the truth. We choose this
neighborhood so that an action with loss at least $\bar\ell$
at the truth has loss at least $\bar\ell/2$ throughout it.
We divide play into epochs of at most
$\lfloor H\rfloor$ periods, where $H=(1-\beta)^{-1}$ is the optimization horizon. At the start of each
epoch, we record the cumulative count $n$ of each source--outcome
pair, and end the epoch early  if any pair is observed $n+1$
additional times. These thresholds are fixed within an epoch
and recomputed at the start of the next; in the first epoch,
all counts are zero. Ending early ensures that no pair is observed
more than $n+1$ additional times within an epoch, as the likelihood
comparison below requires. Since an epoch lasts at most
$\lfloor H\rfloor$ periods and $\beta^{\lfloor H\rfloor}\ge1/4$
for $H\ge2$, each costly action contributes at least
$\bar\ell/8$ to discounted regret at every parameter in
the neighborhood. Integrating this inequality over the neighborhood
under the posterior and applying the Bayesian regret bound
bounds the corresponding average number of costly actions.

To transfer the regret bound to the true parameter, we compare
the probabilities of observations within an epoch under the
true parameter and under the updated Dirichlet benchmark.
With uniformly high probability, the benchmark assigns
probabilities close to the truth for sources with many past
observations. Replacing true draws from these sources by draws
according to the benchmark changes the expected number of
costly actions in an epoch by only a bounded amount. For sources
with fewer past observations, a likelihood comparison bounds
the ratio of true to benchmark probabilities. Only these sources
require an empirical-divergence bound, so its failure probability
can be made small at all ages, independently of the evaluation
horizon. Our prior restriction also bounds the ratio of the
probabilities assigned to the same observations by the benchmark
and the agent's actual posterior, following
\citet[Lemma~A.2]{Fudenberg1993}. The posterior-mass bound enters
at this step: if the posterior at the end of an epoch still assigns
mass at least $\kappa$ to the neighborhood of the truth, Bayes' rule bounds the
posterior predictive probability of the epoch's observations by
$1/\kappa$ times the part contributed by parameters in the
neighborhood, which the integrated bound controls. Together these comparisons
transfer the posterior regret bound to the true parameter.
Summing over epochs gives the result for every $\lambda\ge\beta$.
Shorter evaluation horizons follow from the case $\lambda=\beta$:
losses are nonnegative and $\lambda^t\le\beta^t$ for $\lambda<\beta$, so
$R_\lambda\le(1-\lambda)R_\beta/(1-\beta)\le C_0R_\beta$.
Theorem~\ref{thm:learning} also holds when the reference prior is a finite
mixture of products of Dirichlet distributions, provided the
density of $q_0$ relative to this mixture is bounded above and
away from zero. This allows different mixture components to have
different boundary exponents, so the prior need not be comparable
to any single product Dirichlet. Proposition~\ref{prop:mixtures}
in the Online Appendix states and proves the extension, which
also applies to the deterministic-lifetime results below.

\begin{remark}[Relation to the earlier proof]
The proof of Lemma~\ref{lem:bayesregret}, like the proof of
Lemma~5.3 in \citet{Fudenberg1993}, compares optimal behavior
with a policy that experiments for finitely many periods and
then uses the observations to choose an action to play
thereafter. Patience makes the initial experimentation cost
small relative to the subsequent payoff gain.
Lemma~\ref{lem:mass} gives a pathwise version of
Lemma~5.6 in that paper, controlling posterior mass near
the truth simultaneously in all periods. The earlier proof
makes lifetimes sufficiently long at each fixed discount
factor that further information has negligible value for
most agents. We instead bound regret over short epochs.
Lemmas~\ref{lem:kl}, \ref{lem:ratio},
and~\ref{lem:epochcomparison} let us convert the regret
bound averaged over the posterior into a bound at each
true parameter. This comparison holds uniformly over the
age at which an epoch begins, so we need no restriction
on how quickly lifetimes grow relative to patience.
\hfill$\diamond$
\end{remark}

An immediate implication of Theorem~\ref{thm:learning} is that in a multi-arm bandit, the agent's expected discounted payoff (normalized by the optimization horizon)  converges to the full-information payoff uniformly over the true parameter as the agent's patience grows.

\begin{corollary}[Finite-outcome bandits]\label{cor:bandit}
Suppose arm $i$ has a known finite set of payoff outcomes and unknown probability vector $\theta_i$. Under Assumption~\ref{asst:1},  every Bayesian-optimal policy has expected normalized discounted payoff converging to $\max_i u(i,\theta)$ uniformly over $\theta$ as its discount factor tends to one.
\end{corollary}
\begin{proof}
Represent each arm $i$ as an action that reveals one draw from
source $i$. Apply Theorem~\ref{thm:learning} with $\lambda=\beta$.\end{proof}

\section{Convergence to Nash Equilibrium }\label{sec:steady}

Consider a finite extensive-form stage game with perfect recall, possibly including moves by Nature. We describe a steady-state framework where large populations of agents initially uncertain about how their opponents play  learn others' behavior through repeated interactions. This is the same framework that was introduced in \citep{Fudenberg1993}, except we replace their deterministic lifetimes with geometric random lifetimes (as in \citep{fudenberg2018type}). We will show that, as a consequence of Theorem~\ref{thm:learning}, steady-state behavior converges to a Nash equilibrium of the stage game as agents become more patient and long-lived, regardless of how they approach this joint limit in patience and lifespan.

There is a unit mass of agents for each player role in the game. Each period, agents are independently matched across roles, choose pure strategies for the stage game, and observe the terminal node of their own match, including all actions along the path. They receive payoffs based on the outcome of the game. An agent's personal history records the pure strategy she chose
and the terminal node she observed in each previous match;
new agents start with empty histories. Matching and Nature's moves are independent across periods and across matches. Agents survive with probability $\gamma\in[0,1)$ and discount future payoffs conditional on survival by $\delta\in[0,1)$. A mass $1-\gamma$ of new  agents  are added to each population at the start of each period to replace those who exit. The agent survives for $t$ periods with probability $\gamma^t$
and discounts the payoff conditional on survival by $\delta^t$,
so she optimizes with discount factor $\beta=\delta\gamma$. In the stationary
population, the fraction of agents of age $t$ is $(1-\gamma)\gamma^t$.

Each agent believes that opponents use fixed behavioral strategies; they  know Nature's randomization probabilities at some (or possibly none) of Nature's information sets. For each player role $i$, let $\mathcal{I}_{-i}$ be the collection of information sets in the game tree that belong to $i$'s opponents, plus information sets where $i$ is uncertain about Nature's randomization probabilities. Write $A_h$ and $\theta_h\in\Delta(A_h)$ for the moves and behavior at such an information set $h$. For convenience, we suppose that all new agents in player role $i$ start with the same prior belief over $(\theta_h)_{h \in \mathcal{I}_{-i}}$.\footnote{It is straightforward to extend this to finitely many different priors for each player role population.} These  priors are  fixed independently of $\delta$ and $\gamma$ and satisfy  Assumption~\ref{asst:1}. Agents in role $i$ update their beliefs about  $(\theta_h)_{h \in \mathcal{I}_{-i}}$ based on what they observed in their past matches, and choose a policy mapping from personal histories to pure strategies in the extensive-form game as to maximize expected $\beta$-discounted utility. 

A steady state consists of optimal policies and stationary distributions of personal histories. That is, the distribution of personal histories within each population remains constant from one period to the next as agents match, play according to their policies, add to their personal histories, and some agents exit and new agents enter. \citet[Proposition~1]{fudenberg2022CFH} shows that steady states exist. Each steady state induces an aggregate strategy profile $(\pi_i)_i$, where $\pi_i$ refers to the distribution over pure strategies that population $i$ agents will play in their next match. Note that this distribution puts weight $(1-\gamma)\gamma^t$ on age $t$ agents, whereas the policies of the agents are optimal with respect to the discount factor $\beta= \delta \gamma$.

\begin{theorem}[Nash convergence]\label{thm:nash}
Let $\mathcal E(\delta,\gamma)$ denote the set of steady-state aggregate strategy profiles   under $(\delta, \gamma)$. If $\delta_n\to1$ and $\gamma_n\to1$ and $\pi^n \in \mathcal E (\delta_n,\gamma_n)$ for every $n$, then every accumulation point of  $\pi^n$ is a Nash equilibrium. No restriction on the relative rates with which  $\delta$ and $\gamma$ go to $1$ is required: the result holds   along any joint limit of $(\delta,\gamma)$, and along either order of iterated limit.
\end{theorem}
Proposition~\ref{prop:det-nash} in Appendix~\ref{sec:deterministic} gives the same conclusion for deterministic lifetimes, again at any relative rates of $T\to\infty$ and $\delta\to1$.

\begin{proof}[Proof of Theorem~\ref{thm:nash}]
Fix a steady state and a role $i$. By perfect recall and Kuhn's theorem, applied to each opponent role separately, each opponent role's steady-state aggregate strategy  is realization-equivalent to a behavioral strategy. So we may find mixed actions $\theta_h$ for information sets $h \in \mathcal{I}_{-i}$ such that the learning problem of a new role $i$ agent in this steady state is the same as the problem described in Section~\ref{sec:common} with $(\theta_h)_{h \in \mathcal{I}_{-i}}$ as the true parameters, pure strategies as the action set $S$, the  tree induced by fixing $i$'s choices according to $s$
as the observation tree  for any $s\in S$, and $\beta = \delta\gamma$ as the optimization discount factor. By perfect recall, no information set (i.e., source) appears more than once along any path in the game tree. 

The stationary population weights age $t$ by
$(1-\gamma)\gamma^t$, so role $i$'s aggregate payoff loss
relative to a best response is  the regret of its
learning policy evaluated at $\lambda=\gamma$.
Since $\lambda=\gamma\ge\delta\gamma=\beta$,
Theorem~\ref{thm:learning} applies with $C_0=1$.
Along any sequence with $\delta_n,\gamma_n\to1$, the
optimization discount factor $\beta_n=\delta_n\gamma_n$
tends to one. The theorem therefore makes this aggregate
payoff loss vanish, uniformly over the true parameters
induced by the steady states. Continuity of stage-game
payoffs implies that, at every accumulation point
$\pi^*$, $\pi_i^*$ is a best response to $\pi_{-i}^*$.
Since this holds for every role $i$, $\pi^*$ is Nash.
 
To see that the result holds for the iterated limits, consider  the order in which $\gamma\to1$ for fixed $\delta$, and then $\delta\to1$.
Suppose $\delta_n\to1$ and, for each $n$, $\pi^{n,m}\in\mathcal E(\delta_n,\gamma_{n,m})$
with $\gamma_{n,m}\to1$ and $\pi^{n,m}\to\bar\pi^n$ as $m\to\infty$,
and suppose $\bar\pi^n\to\pi$.
Choose $m(n)$ such that $\|\pi^{n,m(n)}-\bar\pi^n\|<1/n$ and $\gamma_{n,m(n)}>1-1/n$,
and set $\rho^n=\pi^{n,m(n)}$ and $\gamma_n=\gamma_{n,m(n)}$.
Then $\rho^n\in\mathcal E(\delta_n,\gamma_n)$, $(\delta_n,\gamma_n)\to(1,1)$,
and $\rho^n\to\pi$, so $\pi$ is Nash by the first part of the proof.
For the other order, with $\delta\to1$ for fixed $\gamma$ and then $\gamma\to1$,
the same construction applies with the roles of $\delta$ and $\gamma$ reversed. 
\end{proof}

\section{Applications: Product Design and  Congestion}\label{sec:bandit-design}

\subsection{Product Design}\label{sec:design}

Following the experience-good approach of \citet{Shapiro1982}, consider a  seller who designs a product's characteristics and posts a price and a buyer who learns the product's characteristics by repeatedly purchasing and consuming it. The seller's profit-maximizing  design and price can vary with 
patience;  we apply our uniform regret bound to establish convergence of the seller's profit to the full-information monopoly benchmark as patience grows. 

At the start of the interaction, the seller chooses the product characteristics $\theta$ from a compact feasible set $\Theta$ (e.g., a machine's defect frequencies and output quality distribution), which  determines the seller's unit cost $c(\theta)\ge 0$ for the product. The choice of $\theta$ is not observed by the buyer.  The seller also publicly posts a constant price $p\in[0,1]$.

After the seller posts the price, each period  the buyer chooses between buying the product (``Buy'') and taking an outside option (``Outside''). Outside pays zero and reveals no information. Buying generates observations along a known finite tree, whose unknown randomization probabilities at non-terminal nodes are the unknown product characteristics.  Each terminal node $z$ of the tree is associated with a known buyer value $v(z)\in[0,1]$.  So, in each period where buyer chooses Buy and reaches the terminal node $z$, the buyer's payoff is $v(z)-p$ and the seller's payoff is $p - c(\theta)$. We assume both the expected value  $\bar v(\theta)=\mathbb E_\theta v(z)$  and the seller's unit cost $c(\theta)$ are continuous in $\theta$. 

The buyer's prior over $\theta$ satisfies  Assumption~\ref{asst:1}. She  discounts at $\beta$ and uses any optimal policy $\sigma\in\Opt_\beta(q_0;g_p)$, where  $g_p$ assigns payoff  $v(z)-p$ to each terminal node $z$ of Buy and $0$ to Outside. The seller discounts at $\lambda\in[0,1)$.  Define $
 \phi_{\lambda}(\theta,p;\sigma)=(1-\lambda)\mathbb E^\sigma_\theta\sum_{t\ge0}\lambda^t\1\{\mathrm{Buy}_t\}$ as the normalized discounted number of times that the buyer will buy and $\Pi_{\lambda}=[p-c(\theta)]\,\phi_{\lambda}$ as the seller's normalized discounted profit, both with the seller discount factor $\lambda$.  
The buyer's normalized regret at evaluation discount factor $\lambda$ is
\[
 R_{\lambda}(\theta,p;\sigma)=[\bar v(\theta)-p]_+(1-\phi_{\lambda})+[p-\bar v(\theta)]_+\phi_{\lambda}.
\]

Let $\mathcal S(\theta)=\bar v(\theta)-c(\theta)$ be social
surplus and assume $\mathcal S^*=\max_{\theta\in\Theta}
\mathcal S(\theta)>0$. An informed buyer purchases at any
price below $\bar v(\theta)$, so the seller's profit supremum
is $\mathcal S^*$. To approach this benchmark with learning,
the seller can choose a surplus-maximizing design and price
slightly below its mean value. Each missed purchase then
imposes a positive loss on the buyer, so the regret bound
implies that the discounted share of missed sales vanishes
as the buyer becomes patient. By setting the price sufficiently
close to the mean value, the seller can earn profit arbitrarily
close to $\mathcal S^*$.

\begin{proposition}[Optimal product design in the patient limit]
\label{prop:design}
For each $\beta$ and $p$, fix an optimal buyer policy $\sigma_\beta(p)\in\Opt_\beta(q_0;g_p)$, and fix
$\lambda_\beta\in[0,1)$ satisfying
$1-\lambda_\beta\le C_0(1-\beta)$.
For every sequence of seller choices $(\theta_\beta,p_\beta)$
whose profits are within $\eta_\beta\to0$ of the supremum,
as $\beta\uparrow1$,
\begin{equation}\label{eq:D3}
 \Pi\to\mathcal S^*,\quad
 \mathcal S(\theta_\beta)\to\mathcal S^*,\quad
 \phi\to1,\quad
 p_\beta-\bar v(\theta_\beta)\to0,\quad
 \operatorname{dist}\bigl(\theta_\beta,\argmax_\Theta\mathcal S\bigr)\to0.
\end{equation}
If the surplus-maximizing design is unique, $\theta_\beta$
converges to it.
\end{proposition}

The proof is in Appendix~\ref{app:design}. Total surplus per purchase is $\mathcal S(\theta_\beta)\to\mathcal S^*$, and the buyer's surplus $(\bar v(\theta_\beta)-p_\beta)\phi$ tends to zero; patient learning therefore lets the seller extract essentially all of the surplus. The argument applies to multidimensional designs without monotonicity restrictions on demand or learning.

\begin{example}[Relative patience and product quality]\label{ex:quality}
\normalfont
First suppose the seller and buyer have the same discount
factor $\lambda=\beta$. Suppose purchases pay $Y-p$ with $Y\sim\mathrm{Bernoulli}(\theta)$.
The seller's unit cost is  $c(\theta)=\theta^2$,  and the buyer's prior over $\theta$ is Dirichlet $(1,1)$.    Then $\bar v(\theta)=\theta$, $\mathcal S(\theta)=\theta-\theta^2$, $\mathcal S^*=1/4$, and the unique socially optimal $\theta$ is $\theta^*=1/2$. By Proposition~\ref{prop:design}, as $\beta\uparrow1$ every sequence of asymptotically optimal choices has $\theta_\beta\to1/2$, $p_\beta\to1/2$, $\phi\to1$, and $\Pi\to1/4$.

Now maintain $\lambda=\beta$,  and consider the myopic limit
$\beta\downarrow0$. For small $\beta>0$, offering quality zero
at price $p=1/2-\beta$ induces an immediate purchase: buying once
and then stopping has positive expected payoff, and waiting
reveals nothing. The first sale alone yields normalized profit
$(1-\beta)(1/2-\beta)$. Conversely, a price that induces any
purchase must satisfy $p\le 1/2+\beta/(1-\beta)$. The buyer expects to gain $1/2-p$ from her initial purchase, and her discounted payoff from all later periods
is at most $\beta/(1-\beta)$, so the seller's profit is at most $1/2+\beta/(1-\beta)-\theta^2$. Thus, as $\beta\downarrow0$, the supremum of the seller's
normalized profit over quality and price tends to $1/2$,
and every asymptotically optimal choice has quality tending to zero.

Now relax the assumption that $\beta=\lambda$, and  fix quality at zero and price at $1/2$. After any fixed
number of failed purchases, the buyer's posterior still assigns
positive probability to $\theta>1/2$, so by Lemma~\ref{lem:bayesregret},
her continuation value is  positive for sufficiently
high $\beta$, and since Outside pays zero and leaves beliefs
unchanged, Buy is  strictly optimal. For each $n\ge2$,
choose $\beta_n$ close enough to one that the buyer makes at
least $n$ consecutive purchases at quality zero, and require
$1-\beta_n\le n^{-2}$. Set $\lambda_n=1-n^{-1/2}$. Then
$(1-\beta_n)/(1-\lambda_n)\to0$, while $\lambda_n^n\to0$.
The seller's normalized profit is between
$(1-\lambda_n^n)/2$ and $1/2$, so it converges to $1/2$,
above the informed monopoly benchmark of $1/4$. Thus the
conclusion of Proposition~\ref{prop:design} can fail when the
seller's horizon is a vanishing fraction of the buyer's.
\hfill$\diamond$ \end{example}

\subsection{Congestion}\label{sec:congestion}
Consider a large
population of travelers who repeatedly commute through a congested
network, and learn the delay distributions on its edges from their own
trips. These distributions depend on the aggregate traffic flow, which
varies with patience; we apply our uniform regret bound to show that
steady-state flows approach Wardrop equilibrium as travelers become
patient and long-lived, at any relative rates. For affine mean delays,  we also bound the welfare loss and show that fixed marginal-cost tolls, which induce efficiency at
Wardrop equilibrium, yield efficiency in the patient learning limit.

\paragraph{Model.}
Travelers move through a finite network of nodes and directed edges. There are finitely many origin--destination classes $k$, with masses $\rho_k>0$, $\sum_k\rho_k=1$. Travelers in class $k$ go from an origin node $o_k$ to a destination node $d_k\neq o_k$, and different classes have different origin--destination pairs. A \emph{route} from node $o$ to node $d$ is the set of edges of a walk from $o$ to $d$ that does not repeat any edge. Let $\mathcal P_k$ be the set of routes from $o_k$ to $d_k$, which we assume is nonempty. Since a route determines its origin and destination, the sets $\mathcal P_k$ are disjoint for different $k$. A route flow $F=(F_r)$ gives the mass $F_r\ge0$ of travelers on each route $r\in\bigcup_k\mathcal P_k$. Let $\mathcal F$ be the set of feasible route flows, that is, distributions of each class's travelers over its routes: $\sum_{r\in\mathcal P_k}F_r=\rho_k$ for every class $k$.  The edge loads of $F$ are $x_e(F)=\sum_{r\ni e}F_r$. Edge $e$ has a finite set of possible delays $A_e\subset[0,\bar w]$ and, at stationary load $x_e$, a delay distribution $\theta_e(x_e)\in\Delta(A_e)$ with mean $\mathsf t_e(x_e)$. The maps $\theta_e$ are continuous. Each traveler's delay draws are independent across periods, edges, and travelers, conditional on the stationary flow. A traveler observes the delay on every edge of her route. Edges are the sources in the learning problem. A known fixed toll $\chi_e\ge0$ is charged on each edge, and $\chi_r=\sum_{e\in r}\chi_e$. A traveler's payoff is minus the delay and toll along her route; write $g^\chi\in\mathcal G_U$ for this payoff assignment.

Travelers have the survival and discount structure of Section~\ref{sec:steady}:  survival $\gamma$, pure discount $\delta$, $\beta=\delta\gamma$, and births replace deaths. New class-$k$ travelers have a fixed prior $q_{k,0}$ over the edge delay distributions satisfying Assumption~\ref{asst:1}. Travelers do not observe the aggregate flow, other travelers' policies, or the congestion technology $\theta_e(\cdot)$. At a stationary flow $F$, the true parameter of their learning problem is
$
 \theta(F)=\bigl(\theta_e(x_e(F))\bigr)_e.$ Given a selection of optimal class policies $\sigma_k\in\Opt_\beta(q_{k,0};g^\chi)$, a \emph{stationary learning equilibrium} is a flow $F\in\mathcal F$ that equals 
the average route choices of the travelers: 
\begin{equation}\label{eq:congestion-fixed-point}
 F_r=\rho_k(1-\gamma)\,\mathbb E^{\sigma_k}_{\theta(F)}\sum_{t\ge0}\gamma^t\1\{s_t=r\},\qquad r\in\mathcal P_k.
\end{equation}

\begin{proposition}[Existence]\label{prop:congestion-existence}
For every $\beta\le\gamma<1$ and every selection of optimal class policies, a stationary learning equilibrium exists.
\end{proposition}

For fixed policies, aggregate route choices depend continuously
on the stationary flow. Existence follows from Brouwer's theorem.
The proofs of Propositions~\ref{prop:congestion-existence},
\ref{prop:wardrop}, and~\ref{prop:welfare} are in Appendix~\ref{app:congestion}.
The calculations for Example~\ref{ex:two-link}
are in Online Appendix~\ref{sec:OLA_cong}.

A \emph{Wardrop equilibrium}  (\citet{Wardrop1952}) is a Nash equilibrium in a
congestion game with a continuum of individually negligible
travelers: all used routes on a given origin-destination pair have the same expected travel time  plus toll, and no unused route would have strictly lower expected total cost.   Let $c_r(F)=\sum_{e\in r}\mathsf t_e(x_e(F))$ be the
mean delay of route $r$, and define the average excess cost
\[
 G_\chi(F)=\sum_k\sum_{r\in\mathcal P_k}F_r\Bigl[c_r(F)+\chi_r-\min_{r'\in\mathcal P_k}\bigl(c_{r'}(F)+\chi_{r'}\bigr)\Bigr],\qquad G=G_0.
\]
This gap is the reduction in average cost available if each
traveler switched to a cheapest route, holding edge loads fixed.
It is zero  at a Wardrop equilibrium.

\begin{proposition}[Convergence to Wardrop equilibrium]\label{prop:wardrop}
Fix tolls $\chi$. There is $\varepsilon(\beta)\to0$ such that every stationary learning equilibrium $F^{\beta,\gamma}$ with $\gamma\ge\beta$ satisfies
$
 0\le G_\chi(F^{\beta,\gamma})\le\varepsilon(\beta).$ Consequently the distance from $F^{\beta,\gamma}$ to the set of tolled Wardrop equilibria tends to zero as $\beta\uparrow1$, along every $\gamma\ge\beta$ and every $\delta=\beta/\gamma$.
\end{proposition}
At a stationary learning equilibrium, $G_\chi(F)$ equals
travelers' average regret, weighted by class size. Theorem~%
\ref{thm:learning} bounds this regret uniformly over the delay
distributions induced by the equilibrium flow.

\paragraph{Welfare.}
For affine mean delays, \citet{RoughgardenTardos2002}
establishes the $4/3$ price-of-anarchy bound: at a Wardrop
equilibrium, total expected travel time is at most $4/3$ times its minimum
over feasible flows. This bound applies to exact equilibria, at
which no traveler can reduce her expected travel time by switching routes.
\citet[Theorem~6.2]{BlumEvenDarLigett2010} extends it to sequences
of flows in which travelers have small average regret. A flow that
stays constant over time has average regret $G(F)$, so their
argument gives \eqref{eq:poa}.
The next proposition applies this bound to our learning model
and bounds the welfare loss under optimal fixed tolls.

Suppose $\mathsf t_e(x)=\mathsf a_ex+\mathsf b_e$ with $\mathsf a_e,\mathsf b_e\ge0$,  
and let $\mathcal C(F)=\sum_ex_e\mathsf t_e(x_e)$ be total expected travel time. Let $F^{SO}$ minimize $\mathcal C$, with edge loads $x^{SO}$ and value $\mathcal C^{SO}$. The optimal toll
$\chi_e^*=\mathsf a_ex_e^{SO}$ equals the congestion cost that an
additional traveler imposes on others at the efficient flow. These tolls make Wardrop equilibria socially optimal.

\begin{proposition}[Welfare]\label{prop:welfare}
For untolled flows,
\begin{equation}\label{eq:poa}
 \mathcal C(F)\le\tfrac43\bigl(\mathcal C^{SO}+G(F)\bigr).
\end{equation}
With the fixed tolls $\chi^*_e=\mathsf a_ex^{SO}_e$, which do not depend on $\beta$,

\begin{equation}\label{eq:tolls}
 0\le\mathcal C(F)-\mathcal C^{SO}\le G_{\chi^*}(F).
\end{equation}
Hence stationary learning equilibria under untolled traffic have total expected travel time at most $\frac43\mathcal C^{SO}+o(1)$ as $\beta\uparrow1$, and under the fixed tolls $\chi^*$ their total expected travel time converges to $\mathcal C^{SO}$.
\end{proposition}

The planner needs to know the slopes $\mathsf a_e$ and the efficient loads $x^{SO}_e$ to compute $\chi^{*}_{e}. $  If she sets some other fixed toll $\chi$ instead, the welfare loss rises by at most twice the largest error in route tolls.\footnote{Note that $\chi$ and $\chi^*$ enter $G_\chi(F)$ both through the toll on each used route and through the cheapest-route benchmark, so 
$G_{\chi^*}(F)\le G_\chi(F)+2\max_r|\chi_r-\chi^*_r|.$}
 By \eqref{eq:tolls}
and Proposition~\ref{prop:wardrop}, stationary learning equilibria under
$\chi$ therefore have total expected travel time at most
$\mathcal C^{SO}+2\max_r|\chi_r-\chi^*_r|+o(1)$ as $\beta\uparrow1$.

\begin{figure}[tbp]
\captionsetup{labelformat=default,labelsep=period}
\centering
\begin{adjustbox}{max width=\textwidth}
\begin{tikzpicture}[font=\small,>=Stealth]
\definecolor{learnblue}{RGB}{24,83,130}
\definecolor{learngreen}{RGB}{30,119,95}
\begin{scope}[x=5.5cm,y=3.8cm]
\node[anchor=west,font=\bfseries\small] at (-.08,1.43) {(a) Route costs};
\draw[->] (0,0)--(1.08,0);
\draw[->] (0,0)--(0,1.34);
\foreach \x/\lab in {0/0,.25/{1/4},.5/{1/2},1/1}
  {\draw (\x,0)--(\x,-.018);\node[below=3pt] at (\x,0) {$\lab$};}
\foreach \y/\lab in {.5/{1/2},1/1}
  {\draw (0,\y)--(-.012,\y);\node[left=3pt] at (0,\y) {$\lab$};}
\node[rotate=90] at (-.18,.72) {Expected cost};
\node at (.5,-.22) {Share on route $A$, $x$};
\draw[gray!65,densely dotted] (.25,0)--(.25,.5);
\draw[gray!65,densely dotted] (.5,0)--(.5,.5);
\draw[learnblue,thick] (0,0)--(1,1);
\draw[learngreen,thick,dashed] (0,.25)--(1,1.25);
\draw[black,thick] (0,.5)--(1,.5);
\node[anchor=west,text=learnblue,fill=white,inner sep=1pt] at (.66,.65) {$A$: $x$};
\node[anchor=west,text=learngreen,fill=white,inner sep=1pt] at (.39,1.12) {$A$ with toll: $x+1/4$};
\node[anchor=west,fill=white,inner sep=1pt] at (.65,.43) {$B$: $1/2$};
\fill[learngreen] (.25,.5) circle[radius=2.5pt];
\fill[learnblue] (.5,.5) circle[radius=2.5pt];
\end{scope}
\begin{scope}[shift={(8cm,0)},x=5.5cm,y=4.7cm]
\node[anchor=west,font=\bfseries\small] at (-.08,1.157) {(b) Total expected travel time};
\draw[->] (0,0)--(1.08,0);
\draw[->] (0,0)--(0,1.08);
\foreach \x/\lab in {0/0,.25/{1/4},.5/{1/2},1/1}
  {\draw (\x,0)--(\x,-.0145);\node[below=3pt] at (\x,0) {$\lab$};}
\foreach \y/\lab in {.5/{1/2},1/1}
  {\draw (0,\y)--(-.012,\y);\node[left=3pt] at (0,\y) {$\lab$};}
\node[rotate=90] at (-.18,.58) {Total expected travel time};
\node at (.5,-.178) {Share on route $A$, $x$};
\draw[gray!65,densely dotted] (.25,0)--(.25,.4375);
\draw[gray!65,densely dotted] (.5,0)--(.5,.5);
\draw[black,thick,domain=0:1,samples=81] plot (\x,{\x*\x+(1-\x)/2});
\node[anchor=west,fill=white,inner sep=1pt] at (.12,.85) {$\mathcal C(x)=x^2+(1-x)/2$};
\fill[learngreen] (.25,.4375) circle[radius=2.5pt];
\fill[learnblue] (.5,.5) circle[radius=2.5pt];
\node[text=learngreen,align=center] (efficient) at (.13,.18) {Efficient\\$7/16$};
\draw[learngreen,->,thin] (efficient.north)--(.25,.42);
\node[text=learnblue,align=center] (untolled) at (.76,.28) {Untolled\\$1/2$};
\draw[learnblue,->,thin] (untolled.north west)--(.51,.485);
\end{scope}
\end{tikzpicture}
\end{adjustbox}
\caption{\textbf{Learning, congestion, and tolls.} In Example~\ref{ex:two-link}, patient stationary flows approach $x=1/2$ without tolls and $x=1/4$ with the fixed toll $1/4$ on route $A$. Panel (a) shows the corresponding intersections of route costs. Panel (b) shows that the toll implements the minimum total expected travel time, $7/16$, compared with $1/2$ without tolls.}
\label{fig:congestion}
\end{figure}
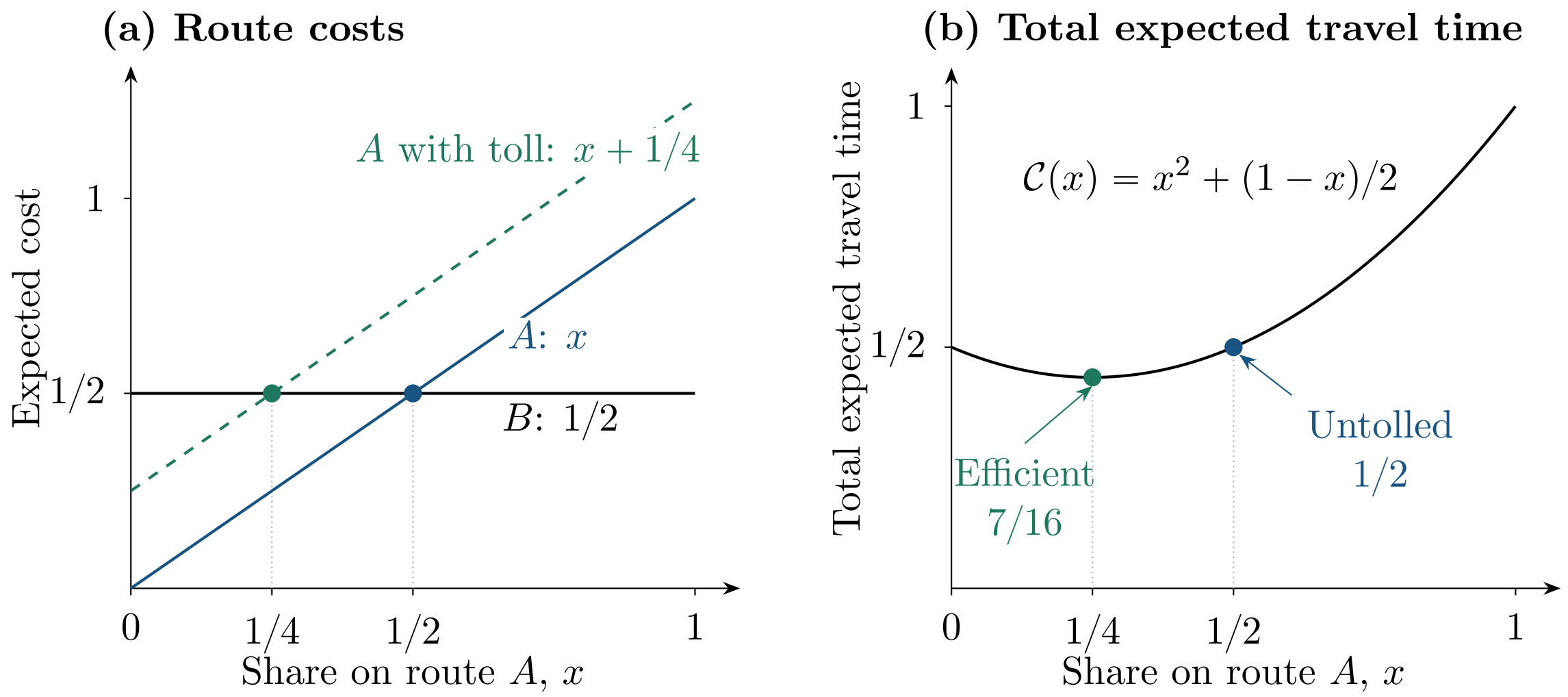

\begin{example}[Two routes]\label{ex:two-link}
\normalfont
A unit mass of identical commuters chooses route $A$ or $B$ each period. Route $A$ has delay $1$ with probability $\theta=x$ and delay $0$ otherwise, where $x$ is the stationary share on $A$. Route $B$ has known delay $1/2$. Every entrant has the prior $\theta\sim\mathrm{Beta}(3,1)=\Dir(3,1)$, with mean $3/4$. Only the chosen route's delay is observed. The Wardrop equilibrium is $x=1/2$.

If $\beta<1/3$, every optimal policy chooses $B$ forever.
The unique stationary flow is therefore $x=0$, where route $A$
would have zero delay. The pessimistic prior prevents travelers
from discovering this. For sufficiently high $\beta$, every
optimal policy starts by trying $A$, so every stationary
equilibrium has positive traffic on that route.

As $\beta\uparrow1$, Proposition~\ref{prop:wardrop} implies
$x_{\beta,\gamma}\to1/2$ uniformly over $\gamma\ge\beta$, where
$x_{\beta,\gamma}$ denotes the share on $A$ in any stationary learning equilibrium.
Thus the true delay probability moves from zero to $1/2$,
with the prior and technology fixed.

Total expected travel time is $\mathcal C(x)=x^2+(1-x)/2$, which is
minimized at $x=1/4$, with value $7/16$. Untolled patient
learning instead leads to $x=1/2$ and total expected travel time $1/2$.
A fixed toll of $1/4$ on route $A$ makes its perceived cost
$x+1/4$, so the patient limit becomes the efficient flow
$x=1/4$. For $\beta<1/2$, however, all travelers still choose
$B$. Figure~\ref{fig:congestion} compares the two patient
limits. Without tolls, travelers learn to equalize private
route costs, but too many use route $A$.
\hfill$\diamond$ \end{example}

\paragraph{Related literature.}
\citet{BlumEvenDarLigett2010} studies convergence to Wardrop
equilibrium under regret-minimizing learning rules, including
rules using only feedback from chosen routes; 
\citet{KricheneDrighesBayen2015} studies related dynamics with
all route costs observed. \citet{WuAminOzdaglar2020} studies
common Bayesian beliefs about a parameter with finitely many
values, updated from aggregate loads and realized costs on all
edges carrying traffic. Loads adjust myopically toward the
allocation preferred under the current belief, and the limit
belief need be correct only about the edges in use. Our
travelers instead choose dynamically optimal policies using only
their own trip histories.

\section{Full Support Is Insufficient}\label{sec:fail}
We now give two examples showing why the prior restriction in Assumption~\ref{asst:1} cannot
be weakened to full support. The first shows that a full-support
prior with an atom can prevent payoff convergence even at a fixed
bandit parameter. The second shows that full-support densities and
independence across opponents do not suffice for Nash convergence
in steady-state learning.

\begin{example}[A full-support prior with an atom]\label{ex:bandit-atom}
\normalfont
Both arms, $A$ and $B$, have reward set $X=\{0,1/2,1\}$, and only the chosen arm's reward is observed.  Let $d_x$ denote the reward distribution concentrated
at $x$, and let $\Theta=\Delta(X)^2$. Set $\theta^0=(d_{1/2},d_0)$,
$\theta^*=(d_{1/2},d_1)$, $\nu=\Dir(1,1,1)\otimes\Dir(1,1,1)$, and
$q_0=\tfrac9{10}\nu^0+\tfrac1{10}\nu$, where $\nu^0$ is the point mass at
$\theta^0$. The prior has full support and correlates the two arms.

At $\theta^*$, arm $A$ pays $1/2$ in every period and arm $B$ pays $1$. Yet for
every $\beta\in(0,1)$, every Bayesian-optimal policy chooses $A$ forever. The
agent can postpone sampling $B$ until a reward from $A$ other than $1/2$ rules
out the pessimistic atom. At $\theta^*$ this never happens, and repeated rewards
of $1/2$ drive the posterior mass on $\theta^0$ to one. Thus even pointwise
convergence fails when Assumption~\ref{asst:1} is replaced by full support.

The failure is not confined to $\theta^*$, which has prior probability zero.
At any $\theta$ near $\theta^*$, arm $A$ pays a reward other than $1/2$ with a
small probability $r(\theta)>0$. Every optimal policy keeps choosing $A$ until
that happens, which takes $1/r(\theta)$ periods on average. Now fix any neighborhood of $\theta^*$ and any $\beta$. The neighborhood contains parameters with $0<r(\theta)\le 1-\beta$.\footnote{ To construct one, give arm $A$ a tiny probability of paying $0$ or $1$, and leave $B$ close to always paying $1$.} At such  parameters, the expected wait $1/r(\theta)$ is at least the optimization horizon $1/(1-\beta)$, so the agent spends most of her effective lifetime on $A$, which pays about $1/2$ while $B$ pays about $1$, so regret  is roughly $1/4$ or more.  Hence regret does not vanish uniformly on any
neighborhood of $\theta^*$, even with $\theta^*$ excluded, although each
neighborhood has positive prior probability.

The argument relies on the atom tying the two arms
together: if the arms were independent under the prior, rewards from $A$ would
reveal nothing about $B$, and there would be nothing to wait for.
The proof is in Online Appendix~\ref{app:bandit-counterexample}.
\hfill $\diamond$
\end{example}

\Needspace{4\baselineskip}

The next example uses a full-support density in a game with additional feedback.

\begin{example}[Non-Nash outcomes with independent beliefs across opponents]\label{ex:1}
\normalfont
The following three-player game, shown in Figure~\ref{fig:counter-game}, has independent beliefs across opponents and a non-Nash steady-state outcome for every patience and lifetime. Let $\bar u=2/5$ be the safe payoff. Player~1 chooses Safe or Risky. Player~3 then chooses Left ($x=0$) or Right ($x=1$)
without observing Player~1's choice. Following Safe, Player~2 chooses Stop or Pass without observing $x$. Risky and Safe--Pass lead to a further choice $y\in\{0,1\}$ by Player~3. At this choice she remembers $x$, but does not distinguish Risky from Safe--Pass. Thus she has two final information sets $h_0,h_1$. Payoffs do not depend directly on $x$.
All players observe the terminal path, including $x$, after each match. The game has perfect recall. Player~2's inability to observe $x$ before choosing is essential.

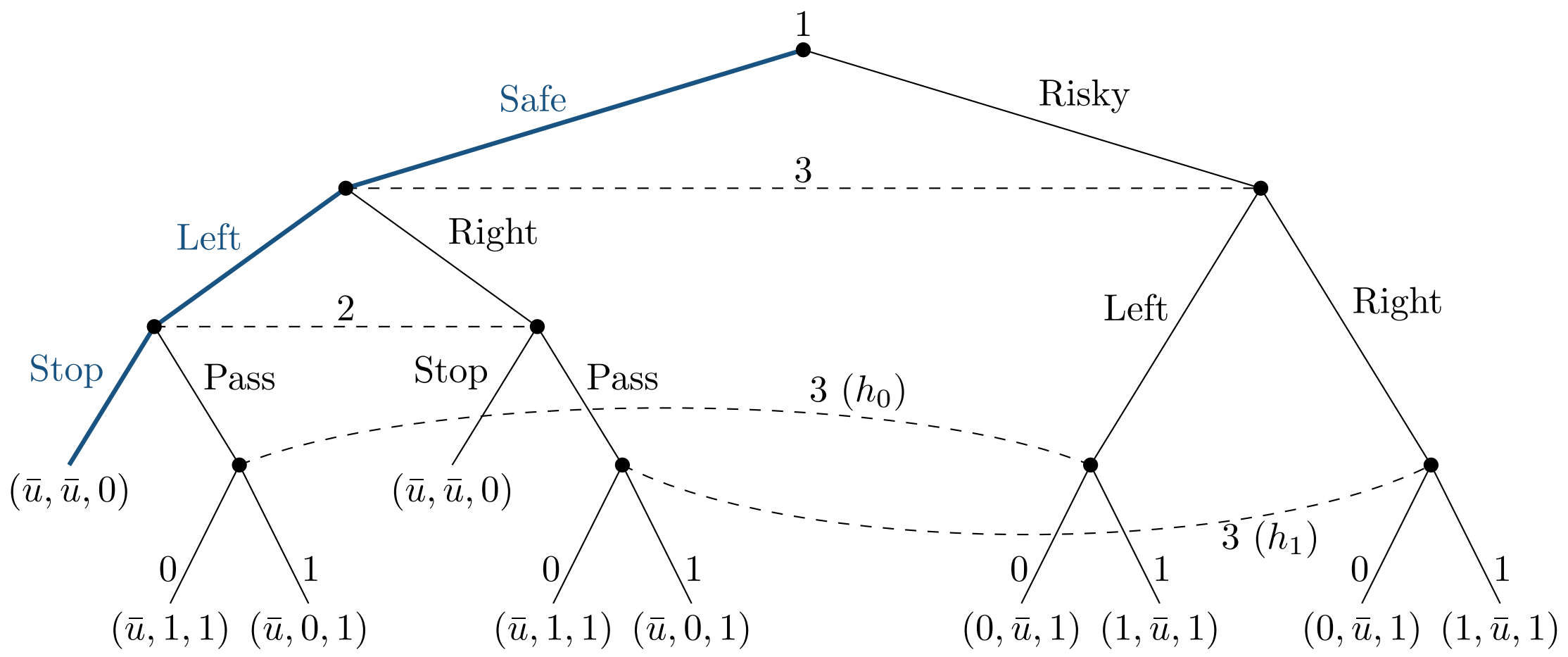
\begin{figure}[tbp]
\captionsetup{labelformat=default,labelsep=period}
\centering
\begin{adjustbox}{max width=\textwidth}
\begin{tikzpicture}[font=\small,x=1cm,y=1.3cm,
  dec/.style={circle,fill,inner sep=0pt,minimum size=4pt},
  lab/.style={font=\footnotesize,inner sep=1.5pt},
  pay/.style={below,font=\footnotesize,inner sep=2pt}]
\definecolor{learnblue}{RGB}{24,83,130}
\coordinate (r) at (0,0);
\coordinate (s) at (-4.3,-1);   \coordinate (k) at (4.3,-1);
\coordinate (sL) at (-6.1,-2);  \coordinate (sR) at (-2.5,-2);
\coordinate (t1) at (-6.9,-3);  \coordinate (sLP) at (-5.3,-3);
\coordinate (t4) at (-3.3,-3);  \coordinate (sRP) at (-1.7,-3);
\coordinate (kL) at (2.7,-3);   \coordinate (kR) at (5.9,-3);
\coordinate (t2) at (-5.95,-4); \coordinate (t3) at (-4.65,-4);
\coordinate (t5) at (-2.35,-4); \coordinate (t6) at (-1.05,-4);
\coordinate (t7) at (2.05,-4);  \coordinate (t8) at (3.35,-4);
\coordinate (t9) at (5.25,-4);  \coordinate (t10) at (6.55,-4);

\draw[dashed] (s)--(k) node[midway,above,lab]{$3$};
\draw[dashed] (sL)--(sR) node[midway,above,lab]{$2$};
\draw[dashed] (sLP) .. controls (-3.5,-2.45) and (0.9,-2.45) .. (kL)
  node[pos=0.7,above,lab]{$3$ ($h_0$)};
\draw[dashed] (sRP) .. controls (0,-3.67) and (4.2,-3.67) .. (kR)
  node[pos=0.77,below,lab]{$3$ ($h_1$)};\draw[learnblue,very thick] (r)--(s) node[midway,above left,lab]{Safe};
\draw (r)--(k) node[midway,above right,lab]{Risky};
\draw[learnblue,very thick] (s)--(sL) node[midway,above left,lab]{Left};
\draw (s)--(sR) node[midway,above right,lab]{Right};
\draw (k)--(kL) node[midway,above left,lab]{Left};
\draw (k)--(kR) node[midway,above right,lab]{Right};
\draw[learnblue,very thick] (sL)--(t1) node[midway,above left,lab]{Stop};
\draw (sL)--(sLP) node[midway,above right,lab]{Pass};
\draw (sR)--(t4) node[midway,above left,lab]{Stop};
\draw (sR)--(sRP) node[midway,above right,lab]{Pass};
\foreach \n/\a/\b in {sLP/t2/t3,sRP/t5/t6,kL/t7/t8,kR/t9/t10}{
  \draw (\n)--(\a) node[pos=0.75,left=1pt,lab]{$0$};
  \draw (\n)--(\b) node[pos=0.75,right=1pt,lab]{$1$};}
\foreach \n in {r,s,k,sL,sR,sLP,sRP,kL,kR} \node[dec] at (\n) {};
\node[above=2pt,lab] at (r) {$1$};
\node[pay] at (t1) {$(\bar u,\bar u,0)$};
\node[pay] at (t4) {$(\bar u,\bar u,0)$};
\foreach \p in {t2,t5} \node[pay] at (\p) {$(\bar u,1,1)$};
\foreach \p in {t3,t6} \node[pay] at (\p) {$(\bar u,0,1)$};
\foreach \p in {t7,t9} \node[pay] at (\p) {$(0,\bar u,1)$};
\foreach \p in {t8,t10} \node[pay] at (\p) {$(1,\bar u,1)$};
\end{tikzpicture}
\end{adjustbox}
\caption{\textbf{The game in Example~\ref{ex:1}.} Dashed lines join nodes in the same information set; the curved ones are Player~3's final information sets $h_0$ and $h_1$, where she chooses $y\in\{0,1\}$. Payoffs are $(u_1,u_2,u_3)$, with $\bar u=2/5$. The thick path is the steady-state outcome (Safe, Left, Stop).}
\label{fig:counter-game}
\end{figure}

Write $p_{\mathrm R}$ for the probability of Risky, $p_{\mathrm P}$ for the probability of Pass, $\xi$ for the probability of $x=1$, and $p_0,p_1$ for the probabilities of $y=1$ at $h_0,h_1$. Conditional on either Risky or Safe--Pass, the mean of $y$ is $P=(1-\xi)p_0+\xi p_1$.

Let $f_{\mathrm{low}}(z)=10(1-5z)_+$ on $[0,1]$. This probability density has support $[0,1/5]$ and mean $1/15$. Player~1 believes $p_{\mathrm P}\sim\operatorname{Beta}(2,1)$ independently of Player~3's strategy, whose density is $\Lambda_+$. Player~2 believes $p_{\mathrm R}\sim\operatorname{Beta}(2,1)$ independently of Player~3's strategy, whose density is $\Lambda_-$, where
\begin{equation}\label{eq:counter-prior}
 \begin{aligned}
 \Lambda_+(\xi,p_0,p_1)&=(1-\xi/2)f_{\mathrm{low}}(p_0)f_{\mathrm{low}}(p_1)+\xi/2,\\
 \Lambda_-(\xi,p_0,p_1)&=(1-\xi/2)f_{\mathrm{low}}(1-p_0)f_{\mathrm{low}}(1-p_1)+\xi/2.
 \end{aligned}
\end{equation}
Player~3 has independent uniform priors over $p_{\mathrm R},p_{\mathrm P}$. All priors are fixed, have continuous bounded densities on the closed parameter spaces, and are strictly positive in the interiors.  These priors violate Assumption~\ref{asst:1} although they satisfy  independence
across opponents and Condition~$\mathcal P$ for every
marginal, as Appendix~\ref{app:counterexample} verifies. 

Importantly, Player~3's first move supplies an observation even when the match
ends at Safe--Stop, and observing Left makes
Player~1 more pessimistic about Risky and Player~2 more pessimistic
about Pass, even though neither observes $y$.

For every $(\delta,\gamma)\in(0,1)^2$, there is a steady state with $p_{\mathrm R}=p_{\mathrm P}=\xi=0$ and $p_0=p_1=1/2$. Every match ends at (Safe, Left, Stop), with payoff vector $(2/5,2/5,0)$. The same outcome is a steady state for every deterministic lifetime and discount factor. Appendix~\ref{app:counterexample} proves optimality; Players~1 and~2 strictly prefer their prescribed choices at histories arising in the steady state, while Player~3 is indifferent among her plans. Each player believes she can learn about $y$ without sacrificing
the safe payoff: Player~1 by waiting for Pass, and Player~2 by waiting
for Risky. The proof shows that waiting is strictly better than
experimenting at every history in the proposed steady state,
regardless of patience or remaining lifetime. Both players therefore
wait, and neither observes $y$.

To see why the outcome is not Nash, suppose a Nash equilibrium assigns positive probability to Safe--Stop. Then $p_{\mathrm R}<1$ and $p_{\mathrm P}<1$. Player~1's use of Safe requires $P\le\bar u$, while Player~2's use of Stop at a reached information set requires $1-P\le\bar u$. These inequalities contradict $\bar u<1/2$. Hence every Nash equilibrium reaches Player~3's final choice with probability one and gives her payoff one. Neither the proposed outcome nor its payoff vector can arise in Nash equilibrium.

Figure~\ref{fig:prior-dependence} illustrates Player~1's growing
pessimism. Under independent priors with the same one-dimensional
marginals, observing $x$ would not change beliefs about $p_0$.

\begin{figure}[t]
\captionsetup{labelformat=default,labelsep=period}
\centering
\begin{adjustbox}{max width=\textwidth}
\begin{tikzpicture}[font=\small,>=Stealth]
\definecolor{learnblue}{RGB}{24,83,130}
\begin{scope}[x=.275cm,y=8cm]
\node[anchor=west,font=\bfseries\small] at (-1.6,.68) {(a) Player 1's expected payoffs};
\draw[->] (0,0)--(21.5,0);
\draw[->] (0,0)--(0,.615);
\foreach \x in {0,5,10,15,20}
  {\draw (\x,0)--(\x,-.009);\node[below=3pt] at (\x,0) {$\x$};}
\foreach \y in {.2,.4,.6}
  {\draw (0,\y)--(-.25,\y);\node[left=3pt] at (0,\y) {$\y$};}
\node[rotate=90] at (-3.6,.32) {Expected stage payoff};
\node at (10,-.105) {Past matches, $n$};
\draw[gray!80,thick,dashed] (0,.5)--(20,.5);
\node[anchor=west,fill=white,inner sep=1pt,text=gray!80] at (7,.535) {True Risky payoff: $1/2$};
\draw[black,thick] (0,.4)--(20,.4);
\node[anchor=west,fill=white,inner sep=1pt] at (10,.435) {Safe payoff: $2/5$};
\draw[learnblue,thick,domain=0:20,samples=101] plot (\x,{1/15+13/(60*(\x+2))});
\fill[learnblue] (0,.175) circle[radius=2pt];
\node[anchor=west,text=learnblue] at (5,.185) {Subjective Risky payoff};
\draw[learnblue,thin,->] (11,.16)--(11,{1/15+13/(60*13)+.009});
\end{scope}
\begin{scope}[shift={(8cm,0)},x=.275cm,y=26.6667cm]
\node[anchor=west,font=\bfseries\small] at (-1.6,.204) {(b) Beliefs about unobserved behavior};
\draw[->] (0,0)--(21.5,0);
\draw[->] (0,0)--(0,.1845);
\foreach \x in {0,5,10,15,20}
  {\draw (\x,0)--(\x,-.0027);\node[below=3pt] at (\x,0) {$\x$};}
\foreach \y in {.05,.10,.15}
  {\draw (0,\y)--(-.25,\y);\node[left=3pt] at (0,\y) {$\y$};}
\node[rotate=90] at (-3.8,.096) {Posterior probability of $p_0>2/5$};
\node at (10,-.0315) {Past matches, $n$};
\draw[gray!80,thick,dashed] (0,.15)--(20,.15);
\node[anchor=west,text=gray!80] at (5,.161) {Independent prior};
\draw[learnblue,thick,domain=0:20,samples=101] plot (\x,{3/(10*(\x+2))});
\fill[learnblue] (0,.15) circle[radius=2pt];
\node[anchor=west,text=learnblue] at (7,.075) {Example's prior};
\draw[learnblue,thin,->] (11,.066)--(11,{3/(10*13)+.003});
\end{scope}
\end{tikzpicture}
\end{adjustbox}
\caption{\textbf{Observations can discourage experimentation elsewhere.} Posteriors after $n$ (Safe, Left, Stop) matches in Example~\ref{ex:1}. Although $y$ is never observed, Player~1's expected Risky payoff approaches $1/15$ (panel a), and her posterior probability that $p_0>2/5$ approaches zero (panel b). Every finite-history posterior retains full support. The dashed line in panel (b) uses the product of the same one-dimensional prior marginals, conditional on the same histories.}
\label{fig:prior-dependence}
\end{figure}
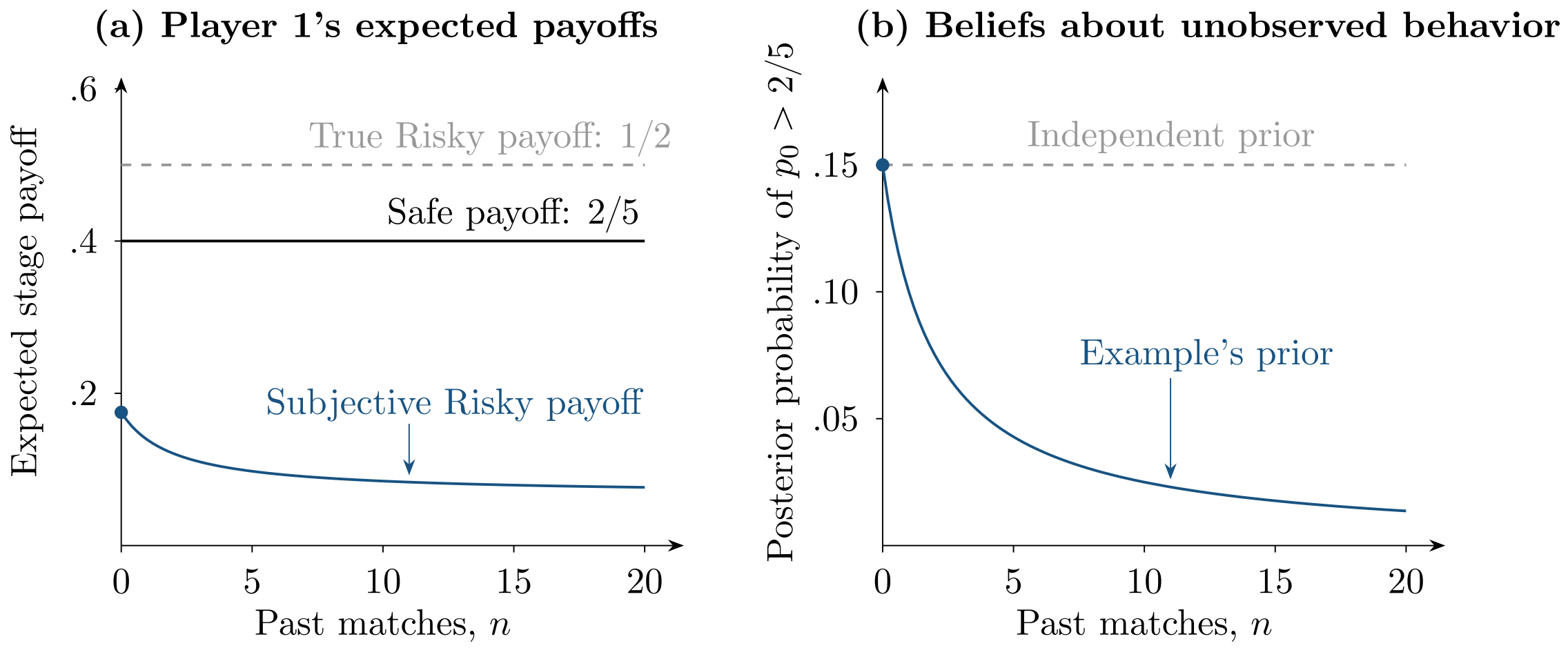\hfill$\diamond$ \end{example}

Example~\ref{ex:1} relies on correlated beliefs across information sets
of the same opponent.
\citet{fudenberg2006superstition} rules this out by requiring independence
across information sets, but allows marginal densities that approach zero at the
boundary faster than any Dirichlet density, which our assumption
excludes.

\citet[Example~1]{LiPan2026} uses dependence of this kind for a
different purpose. Under a singular prior that perfectly correlates
beliefs about one opponent's behavior at two decision nodes, a profile
that is not path-equivalent to any subgame-confirmed equilibrium is a
steady state for every lifetime and discount factor. That profile is a
Nash equilibrium, whereas the priors in Example~\ref{ex:1} have
densities and its outcome is not Nash.

Example~\ref{ex:1}  identifies a problem with the prior
assumption in \citet{fudenberg2022CFH}, because its  priors satisfy
the strict positivity requirement in Section~4 of that paper. 
The example therefore
contradicts that paper's assertion, following Definition~1,
that patiently stable profiles are Nash equilibria when
agents observe terminal nodes. It also disproves the
conjecture in the following paragraph that patient stability
implies Nash equilibrium whenever each player's payoff is
measurable with respect to her terminal-node partition.\footnote{By Theorem~\ref{thm:nash}, the assertion is correct for priors
satisfying Assumption~\ref{asst:1}, so the steps of that paper that rely
on it (the auxiliary-game arguments in its Appendices~A.2 and~A.3 and
one step in the proof of its Proposition~2) go through for such priors.}

\section{Concluding Discussion}

Several questions remain open. Does joint-limit Nash convergence hold for
full-support priors that are independent across all sources, and does uniform
payoff convergence hold in ordinary bandits for correlated full-support
densities? Does it hold for independent priors whose marginal densities vanish
faster than any Dirichlet density at the boundary, which our assumption excludes
but \citet{fudenberg2006superstition} allow? Our bounds also do not give a sharp rate
of uniform convergence, but such results seem possible.  Finally,  even when  all joint limits of steady-states profiles as patience and lifespan grow are Nash equilibria, just which Nash equilibria are limits can depend on the way the double limit is taken. It would be interesting to explore how the relative rates at which $\delta$ and $\gamma$ go to $1$ influence which steady states are accumulation points, but that is beyond the scope of this paper.
\par 
\setstretch{1}
 \setlength{\bibitemsep}{\baselineskip}
 \printbibliography

@article{Shapiro1982,
  author = {Shapiro, Carl},
  title = {Consumer Information, Product Quality, and Seller Reputation},
  journal = {Bell Journal of Economics},
  year = {1982}, volume = {13}, number = {1}, pages = {20--35}}

@article{RoughgardenTardos2002,
  author = {Roughgarden, Tim and Tardos, {\'E}va},
  title = {How Bad Is Selfish Routing?},
  journal = {Journal of the ACM},
  year = {2002}, volume = {49}, number = {2}, pages = {236--259}}

@book{BeckmannMcGuireWinsten1956,
  author = {Beckmann, Martin and McGuire, C. B. and Winsten, C. B.},
  title = {Studies in the Economics of Transportation},
  publisher = {Yale University Press}, year = {1956}}

@inproceedings{WuAminOzdaglar2020,
  author = {Wu, Manxi and Amin, Saurabh and Ozdaglar, Asuman},
  title = {Bayesian Learning with Adaptive Load Allocation Strategies},
  booktitle = {Proceedings of the 2nd Conference on Learning for Dynamics and Control},
  series = {Proceedings of Machine Learning Research},
  year = {2020}, volume = {120}, pages = {561--570}}

@article{Gittins1979,
  author = {Gittins, J. C.},
  title = {Bandit Processes and Dynamic Allocation Indices},
  journal = {Journal of the Royal Statistical Society, Series B},
  year = {1979}, volume = {41}, number = {2}, pages = {148--164}}

@article{
Fudenberg1993,
author = {Fudenberg, Drew and David K. Levine},
title = {Steady state learning and Nash equilibrium},
journal = {Econometrica},
year = 1993,
volume = 61,
pages = {547-573},
}

@article{
fudenberg2006superstition,
author = {Fudenberg, Drew and David K. Levine},
title = {Superstition and rational learning},
journal = {American Economic Review},
year = 2006,
volume = 96,
pages = {630-651},
}

@article{fudenberg2018type,
  title={Learning and type compatibility in signaling games},
  author={Fudenberg, Drew and He, Kevin},
  journal={Econometrica},
  volume={86},
  number={4},
  pages={1215--1255},
  year={2018},
  publisher={Wiley Online Library}
}

@article{diaconis1990uniform,
  title={On the uniform consistency of Bayes estimates for multinomial probabilities},
  author={Diaconis, Persi and Freedman, David},
  journal={The Annals of Statistics},
  volume={18},
  number={3},
  pages={1317--1327},
  year={1990},
  publisher={JSTOR}
}

@article{frostig2016four,
  title={Four proofs of Gittins’ multiarmed bandit theorem},
  author={Frostig, Esther and Weiss, Gideon},
  journal={Annals of Operations Research},
  volume={241},
  number={1},
  pages={127--165},
  year={2016},
  publisher={Springer}
}

@article{Wardrop1952,
  author  = {Wardrop, J. G.},
  title   = {Some Theoretical Aspects of Road Traffic Research},
  journal = {Proceedings of the Institution of Civil Engineers},
  year    = {1952},
  volume  = {1},
  number  = {3},
  pages   = {325--362},
  doi     = {10.1680/ipeds.1952.11259}
}

@article{RussoVanRoy2014,
  author  = {Russo, Daniel and Van Roy, Benjamin},
  title   = {Learning to Optimize via Posterior Sampling},
  journal = {Mathematics of Operations Research},
  year    = {2014},
  volume  = {39},
  number  = {4},
  pages   = {1221--1243},
  doi     = {10.1287/moor.2014.0650}
}

@inproceedings{BubeckLiu2013,
  author    = {Bubeck, S{\'e}bastien and Liu, Che-Yu},
  title     = {Prior-Free and Prior-Dependent Regret Bounds
               for {Thompson} Sampling},
  booktitle = {Advances in Neural Information Processing Systems},
  year      = {2013},
  volume    = {26},
  pages     = {638--646},
  publisher = {Curran Associates, Inc.}
}

@inproceedings{AgrawalGoyal2013,
  author    = {Agrawal, Shipra and Goyal, Navin},
  title     = {Further Optimal Regret Bounds for {Thompson} Sampling},
  booktitle = {Proceedings of the Sixteenth International Conference
               on Artificial Intelligence and Statistics},
  series    = {Proceedings of Machine Learning Research},
  year      = {2013},
  volume    = {31},
  pages     = {99--107},
  publisher = {PMLR}
}

@inproceedings{Lattimore2016,
  author    = {Lattimore, Tor},
  title     = {Regret Analysis of the Finite-Horizon {Gittins}
               Index Strategy for Multi-Armed Bandits},
  booktitle = {29th Annual Conference on Learning Theory},
  series    = {Proceedings of Machine Learning Research},
  year      = {2016},
  volume    = {49},
  pages     = {1214--1245},
  publisher = {PMLR}
}

@article{ChenWangYuanWang2016,
  author  = {Chen, Wei and Wang, Yajun and Yuan, Yang
             and Wang, Qinshi},
  title   = {Combinatorial Multi-Armed Bandit and Its Extension
             to Probabilistically Triggered Arms},
  journal = {Journal of Machine Learning Research},
  year    = {2016},
  volume  = {17},
  number  = {50},
  pages   = {1--33}
}

@inproceedings{WangChen2017,
  author    = {Wang, Qinshi and Chen, Wei},
  title     = {Improving Regret Bounds for Combinatorial
               Semi-Bandits with Probabilistically Triggered
               Arms and Its Applications},
  booktitle = {Advances in Neural Information Processing Systems},
  year      = {2017},
  volume    = {30},
  pages     = {1161--1171},
  publisher = {Curran Associates, Inc.}
}

@article{aghion1991optimal,
  title={Optimal learning by experimentation},
  author={Aghion, Philippe and Bolton, Patrick and Harris, Christopher and Jullien, Bruno},
  journal={Review of Economic Studies},
  volume={58},
  number={4},
  pages={621--654},
  year={1991},
  publisher={Wiley-Blackwell}
}

@article{BlumEvenDarLigett2010,
  author  = {Blum, Avrim and Even-Dar, Eyal and Ligett, Katrina},
  title   = {Routing Without Regret: On Convergence to {Nash}
             Equilibria of Regret-Minimizing Algorithms in Routing Games},
  journal = {Theory of Computing},
  year    = {2010},
  volume  = {6},
  pages   = {179--199},
  doi     = {10.4086/toc.2010.v006a008}
}

@article{KricheneDrighesBayen2015,
  author  = {Krichene, Walid and Drigh{\`e}s, Benjamin and Bayen, Alexandre M.},
  title   = {Online Learning of {Nash} Equilibria in Congestion Games},
  journal = {SIAM Journal on Control and Optimization},
  year    = {2015},
  volume  = {53},
  number  = {2},
  pages   = {1056--1081},
  doi     = {10.1137/140980685}
}

@article{fudenberg2023pathwise,
  title={Pathwise concentration bounds for Bayesian beliefs},
  author={Fudenberg, Drew and Lanzani, Giacomo and Strack, Philipp},
  journal={Theoretical Economics},
  volume={18},
  number={4},
  pages={1585--1622},
  year={2023},
  publisher={Wiley Online Library}
}

@article{fudenberg2022CFH,
title = {Observability, dominance,  and induction in learning models},
author={Clark, D and Fudenberg, D and He, K},
year= {2022},
journal= {Journal of Economic Theory},
volume= {206},
pages= {105569}
}

@article{rothschild1974pricing,
  author  = {Rothschild, Michael},
  title   = {A Two-Armed Bandit Theory of Market Pricing},
  journal = {Journal of Economic Theory},
  year    = {1974},
  volume  = {9},
  number  = {2},
  pages   = {185--202},
  doi     = {10.1016/0022-0531(74)90066-0}
}

@article{fudenberg2017rare,
  author  = {Fudenberg, Drew and He, Kevin and Imhof, Lorens A.},
  title   = {{Bayesian} Posteriors for Arbitrarily Rare Events},
  journal = {Proceedings of the National Academy of Sciences},
  year    = {2017},
  volume  = {114},
  number  = {19},
  pages   = {4925--4929},
  doi     = {10.1073/pnas.1618780114}
}

@article{LiPan2026,
  author  = {Li, Zihao and Pan, Minghao},
  title   = {Rational Learning One Step off the Path},
  journal = {arXiv preprint arXiv:2608.05380},
  year    = {2026}
}
 \onehalfspacing
\appendix
 \section{Proof of Theorem~\ref{thm:learning}}\label{sec:proof}
This appendix proves Theorem~\ref{thm:learning}. Lemmas~\ref{lem:lipschitz}--\ref{lem:epochcomparison} give the payoff, Bayesian-regret, and posterior bounds, and the final subsection combines them to bound regret at each true parameter.

\subsection{Notation and posterior comparison}
We first prove the result for $\lambda\ge\beta$, then extend
it to shorter evaluation horizons at the end of the proof. Write
$b=1-\beta$, $H=b^{-1}$, $\psi=1-\lambda\le b$,
and $K=c_+/c_-$, and restrict attention to $H\ge2$.
The optimization and evaluation horizons are $H$ and
$1/\psi$, respectively. Throughout the proofs, $C$
denotes a finite positive constant that may change between
occurrences. Unless indicated by subscripts, constants depend
only on the observation trees, prior, and payoff bound $U$.

Let $n_{ha}(y)$ count
observations of outcome $a$ from source $h$, and let
$N_h(y)=\sum_a n_{ha}(y)$. Denote the posterior by $q_y$
and the reference posterior by $D_y$. For a fixed history, the probabilities contributed by the
agent's action choices and by known randomization are the
same under every candidate parameter. These factors cancel
from Bayes' rule, leaving the source likelihood
$\prod_{h,a}\theta_{ha}^{n_{ha}(y)}$. Hence $D_y$ is the product Dirichlet below, and $dq_y/dD_y=f/\int f\,dD_y$ lies in $[K^{-1},K]$:
\begin{equation}\label{eq:posterior}
 D_y=\bigotimes_h\Dir(\alpha_h+n_h(y)),\qquad
 K^{-1}\le \frac{dq_y}{dD_y}\le K.
\end{equation}
Optimal policies exist because there are finitely many actions
and payoffs are bounded and discounted.
Every finite history that has positive probability
under an optimal policy and some true parameter has positive prior predictive
probability under that policy, so optimal policies have optimal continuations
at these histories, and every action chosen with positive probability is optimal at that history.

\subsection{Payoff and Bayesian regret bounds}
\begin{lemma}[Payoff continuity]\label{lem:lipschitz}
For every $g\in\mathcal G_U$, action $s$, and parameters $\theta,\vartheta$,
\begin{equation}\label{eq:lipschitz}
 |u_g(s,\theta)-u_g(s,\vartheta)|\le U\sum_h\norm{\theta_h-\vartheta_h}.
\end{equation}
Consequently, if $\max_h\norm{\theta_h-\vartheta_h}<\eta$ with $\eta=\bar\ell/(4MU)$, then $\ell_g(s,\theta)\ge\bar\ell$ implies $\ell_g(s,\vartheta)\ge\bar\ell/2$.
\end{lemma}
\begin{proof}
Couple the two path distributions node by node, using the same known randomization and a maximal coupling of the draws from $\theta_h$ and $\vartheta_h$ whenever source $h$ is used. The paths coincide until the first disagreement, and each source is used at most once on a path, so the paths differ with probability at most $\sum_h\TV(\theta_h,\vartheta_h)=\frac12\sum_h\norm{\theta_h-\vartheta_h}$. Payoffs lie in $[-U,U]$, which gives \eqref{eq:lipschitz}. The maximum $u_g^*$ satisfies the same bound, so $|\ell_g(s,\theta)-\ell_g(s,\vartheta)|\le 2UM\eta=\bar\ell/2$.
\end{proof}

\begin{lemma}[Bayesian regret]\label{lem:bayesregret}
For every posterior $q$ and optimal continuation $\sigma$,
\begin{equation}\label{eq:bayesregret}
 \int\mathbb{E}_\vartheta^\sigma\!\left[\sum_{t\ge0}\beta^t\ell(s_t,\vartheta)\right]q(d\vartheta)
 \le C H^{2/3}.
\end{equation}
Equivalently, the optimal normalized value satisfies
$V_\beta(q)\ge\int u^*\,dq-Cb^{1/3}$.
\end{lemma}
\begin{proof}
Consider the comparison policy that plays each action $m$ times and then always plays an action $\widehat s$ with the highest sample mean payoff. Each play of an action $s$ restarts its observation tree and pays the known payoff at the terminal node reached. Conditional on the parameter $\vartheta$, its realized payoffs are therefore independent draws from a distribution on $[-U,U]$ with mean $u(s,\vartheta)$.

Let $\widehat u_s$ be the mean of the $m$ draws. If $s^*$ maximizes $u(\cdot,\vartheta)$, then $\widehat u_{s^*}\le\widehat u_{\widehat s}$, so $\ell(\widehat s,\vartheta)\le2\max_s|\widehat u_s-u(s,\vartheta)|$. Bounded payoffs thus give expected exploitation loss at most $C/\sqrt m$ per period, uniformly in the parameter, and the exploitation periods have total discount weight at most $H$.

Exploration costs at most $Cm$, so with $m=\lceil H^{2/3}\rceil$ the comparison policy has discounted regret at most $CH^{2/3}$ at every parameter, and hence also on average under $q$. An optimal continuation has at least as high a $q$-expected discounted payoff, and the benchmark $H\int u^*\,dq$ does not depend on the policy, so its integrated regret is no larger. This proves \eqref{eq:bayesregret}.\end{proof}

\subsection{Posterior bounds}
For $\eta>0$, put
$B_\eta(\theta)=\{\vartheta:\max_h\norm{\vartheta_h-\theta_h}<\eta\}$.

\begin{lemma}[Posterior mass near the truth]\label{lem:mass}
For every $\eta,\zeta>0$, there is $\kappa>0$ such that, uniformly over true parameters and policies,
\begin{equation}\label{eq:mass}
 \mathbb{P}_\theta\{q_{y_t}(B_\eta(\theta))\ge\kappa\text{ for all }t\}\ge1-\zeta.
\end{equation}
\end{lemma}
\begin{proof}
Pregenerate independent infinite streams for all sources and reveal the next draw whenever a source is visited.\footnote{Conditional on the true parameter, draw an independent
infinite stream for each source and reveal its next unused
outcome at each visit. This preserves the observation law and
lets us bound empirical errors uniformly over sampling policies.
The proof of Lemma~B.2 in \citet{Fudenberg1993} uses the same
construction, indexing observations by visits to information
sets. For a related use in proofs of the Gittins index theorem,
see \citet[Section~3.2]{frostig2016four}.} Let $\widehat\theta_h(k)$ be the empirical vector of its first $k$ draws. Hoeffding's inequality and a union bound over sources, coordinates, and $k\ge k_0$ imply, with probability at least $1-\zeta$, that all these vectors are within $\eta/4$ of the truth, for a sufficiently large $k_0$ uniform in $\theta$.

Dirichlet mean and variance formulas give $\pE{D_y}{\norm{\vartheta_h-\widehat\theta_h(k)}}\le C/\sqrt{k+1}$. Our bounded-density comparison transfers this bound to $q_y$,
up to a constant factor. Increasing $k_0$, the reference marginal assigns at least $1/2$ to the $\eta$-neighborhood of $\theta_h$ whenever $k\ge k_0$. For $k<k_0$, there are only finitely many count vectors. Choose a finite $\eta/2$-net of each source simplex. Every open $\eta$-ball about a parameter contains an open $\eta/2$-ball about some net point. Each such smaller ball has positive mass under every associated Dirichlet posterior. Taking the minimum over sources, count vectors, and net points gives a common $\rho>0$. Product structure and \eqref{eq:posterior} now give \eqref{eq:mass} with $\kappa=K^{-1}\min\{\rho,1/2\}^M$.
\end{proof}

The next two lemmas control empirical divergences and use them
to compare true and reference predictive probabilities for a single source.

\begin{lemma}[Empirical divergence]\label{lem:kl}
For an iid categorical stream with $d$ outcomes, true vector $p$, and empirical vector $\widehat p_n$, for $x>0$ and integer $T\ge1$,
\begin{equation}\label{eq:kl}
 \mathbb{P}_p\{\exists n\le T:n\KL(\widehat p_n\Vert p)>x\}
 \le 2d(1+\lceil\log_2T\rceil)e^{-x/(2d)}.
\end{equation}
\end{lemma}
\begin{proof}
For a Bernoulli stream, write $\kl(\cdot,\cdot)$ for the divergence between Bernoulli distributions with the given means, and consider $N\le n\le2N$. On the upper-tail crossing of $n\kl(\widehat p_n,p)>x$, the empirical mean exceeds $p_+\ge p$ satisfying $\kl(p_+,p)=x/(2N)$. The likelihood-ratio martingale for $p_+$ against $p$ is then at least $e^{x/2}$. Its maximal inequality bounds the crossing probability by $e^{-x/2}$; the lower tail is identical. If the threshold exceeds the attainable divergence, the event is empty; endpoint cases follow by limits, and $p\in\{0,1\}$ is immediate. Sum over dyadic blocks. For categorical observations, $\KL(v\Vert p)\le\sum_{a=1}^d\kl(v_a,p_a)$: the difference is $(d-1)$ times the divergence between the normalized complementary vectors. A union bound over coordinates, with threshold $x/d$, proves \eqref{eq:kl}.
\end{proof}

The bound holds simultaneously over sample sizes. It therefore
applies when the number of draws from a source depends on the
policy and earlier observations.

\begin{lemma}[Predictive comparison for one source]\label{lem:ratio}
Consider a source with $d$ outcomes, true vector $p$, and reference prior $\Dir(\alpha)$,
and put $Z_\alpha(n)=\int\prod_a v_a^{n_a}\Dir(\alpha)(dv)$ and
$\Psi_\alpha(n;p)=\log(\prod_a p_a^{n_a}/Z_\alpha(n))$, with $0^0=1$.
Let $n$ be a count vector possible under $p$, with total $N$ and empirical
vector $\widehat p$. If further counts $m$ satisfy $0\le m_a\le n_a+1$
for every $a$, then
\begin{equation}\label{eq:increment}
 \log\frac{\prod_a p_a^{m_a}}{Z_\alpha(n+m)/Z_\alpha(n)}
 =\Psi_\alpha(n+m;p)-\Psi_\alpha(n;p)
 \le N\KL(\widehat p\Vert p)+C_\alpha,
\end{equation}
where the divergence term is zero when $N=0$.
\end{lemma}
\begin{proof}
Restrict attention to counts possible under $p$.
After counts $n$, the reference predictive probability of a
sequence with additional counts $m$ is
$Z_\alpha(n+m)/Z_\alpha(n)$, which gives the equality in \eqref{eq:increment}.

The Dirichlet normalizing constant is
$Z_\alpha(n)=\dfrac{\Gamma(|\alpha|)}{\Gamma(|\alpha|+N)}
\prod_a\dfrac{\Gamma(\alpha_a+n_a)}{\Gamma(\alpha_a)}$,
where $|\alpha|=\sum_a\alpha_a$.
For each fixed $c>0$, Stirling's bounds give
$\log\Gamma(k+c)=(k+c-\tfrac12)\log(k+1)-k+O_c(1)$,
with the remainder bounded independently of the integer $k\ge0$.
Applying this to the numerator and denominator of $Z_\alpha(n)$,
the linear terms cancel because $\sum_a n_a=N$.
Moreover, boundedness of $k\log(1+1/k)$ implies
$\sum_a n_a\log(n_a+1)-N\log(N+1)
=\sum_a n_a\log(n_a/N)+O(1)$.
Substituting into the definition of $\Psi_\alpha$ therefore gives
\begin{equation}\label{eq:normalizer}
 \Psi_\alpha(n;p)=-N\KL(\widehat p\Vert p)
 +(|\alpha|-\tfrac12)\log(N+1)
 -\sum_a(\alpha_a-\tfrac12)\log(n_a+1)+O_\alpha(1),
\end{equation}
where $\widehat p=n/N$ for $N>0$, and terms multiplied by zero
are interpreted as zero, including the divergence term when $N=0$.
The remainder is bounded by a constant depending only on $\alpha$,
for all count vectors possible under $p$.

If $0\le m_a\le n_a+1$ for every outcome $a$, then $(n_a+m_a+1)/(n_a+1)\in[1,2]$ and $(N+\sum_a m_a+1)/(N+1)\in[1,d+1]$. Subtracting \eqref{eq:normalizer} at $n$ from its value at $n+m$, the logarithmic terms change by a bounded amount. Dropping the nonpositive terminal divergence gives the inequality in \eqref{eq:increment}.
\end{proof}

The restriction $m_a\le n_a+1$ cannot be dropped. If $p=(1,0)$ and
$n=(N,0)$, then for $m=(m_1,0)$ the left side of \eqref{eq:increment} equals
$\sum_{j=0}^{m_1-1}\log\bigl(1+\alpha_2/(\alpha_1+N+j)\bigr)$, which is
unbounded as $m_1/(N+1)\to\infty$ although the empirical divergence is zero.
For this reason, the epochs in the proof of Theorem~\ref{thm:learning} end
as soon as some cell's increment reaches its initial count plus one.

\subsection{Epoch comparison at every age}
Fix a scale $L\ge2$, put $N_L=\lceil L^6\rceil$ and
$x_L=(\log L)/(6M)$, and write
$d_{\max}=\max_h|A_h|$. Call a history $y$ $L$-good if every source
with $N_h\le N_L$ satisfies
$N_h\KL(\widehat\theta_h\Vert\theta_h)\le x_L$, and every
source with $N_h>N_L$ satisfies
$\norm{\widehat\theta_h-\theta_h}\le L^{-2}$; here $N_h=N_h(y)$ and $\widehat\theta_h=\widehat\theta_h(y)$ is the empirical vector of source $h$'s outcomes in $y$.

\begin{lemma}[Good histories at all ages]\label{lem:goodhistories}
There are constants $C,c'>0$ such that, for every $L\ge2$,
true parameter, and policy,
\begin{equation}\label{eq:goodhistories}
 \begin{gathered}
 \mathbb P_\theta\{y_t\text{ is $L$-good for all }t\}
 \ge1-p_L,\\
 p_L=C(1+\log L)L^{-1/(12Md_{\max})}
       +CL^4e^{-c'L^2}\longrightarrow0.
 \end{gathered}
\end{equation}
\end{lemma}
\begin{proof}
Pregenerate the source streams. Apply Lemma~\ref{lem:kl} with
cutoff $N_L$ and threshold $x_L$, and take a union bound over
sources. This bounds the probability of a failed KL condition
by the first term in $p_L$. Hoeffding's inequality and a union
bound over coordinates give
$\mathbb P_\theta\{\norm{\widehat\theta_h(k)-\theta_h}>L^{-2}\}
\le Ce^{-c'k/L^4}$. Summing over sources and all $k>N_L$
gives the second term. These bounds hold simultaneously over
sample sizes, so they apply to the counts generated by any policy.
\end{proof}
At any history $y$, a continuation record $w$ lists the subsequent
actions and observations, possibly ending at a stopping time.
 Let $\mathbb P_\theta(w\mid y)$ denote  its true
conditional probability under the agent's policy, and write 
$Q_y(w)=\int\mathbb P_\vartheta(w\mid y)q_y(d\vartheta)$ for its
predictive probability. Call each source--outcome pair $(h,a)$
a cell. A record is \emph{short} relative to $y$ if it raises
each cell's count by at most $n_{ha}(y)+1$.

For a set $B$ of parameters, let $Q_y^B(w)=\int_B\mathbb P_\vartheta(w\mid y)q_y(d\vartheta)$. Bayes' rule gives $Q_y^B(w)=q_{yw}(B)Q_y(w)$, where $q_{yw}$ is the posterior after $y$ and then $w$. 
\begin{lemma}[Epoch comparison]\label{lem:epochcomparison}
Fix an $L$-good history $y$ and a continuation policy. Suppose
the epoch ends at a stopping time of at most
$\lfloor L\rfloor$ periods and every resulting record $w$
is short relative to $y$.
For every measurable parameter set $B$, $\kappa>0$, and
record function $\Xi$ with $0\le\Xi\le L$,
\begin{equation}\label{eq:epochcomparison}
 \mathbb E_\theta\!\left[
 \Xi(w)\1\{q_{yw}(B)\ge\kappa\}\mid y\right]
 \le C\kappa^{-1}L^{1/6}
       \int_B\mathbb E_\vartheta[\Xi(w)\mid y]q_y(d\vartheta)+C.
\end{equation}
\end{lemma}
\begin{proof}
Call sources with $N_h(y)>N_L$ heavily sampled, keeping this
classification fixed throughout the epoch. Let $\mathcal A_y$ be the
record law that uses reference Dirichlet predictive draws for
these sources and true draws for all others. It uses the same
policy, known randomization, and stopping rule as the original
experiment. Reference predictive draws use each source's counts,
including those added during the epoch.

For a heavily sampled source with $k=N_h(y)$, initial empirical
error is at most $L^{-2}$. At most $\lfloor L\rfloor$ additional
draws occur during the epoch, so its reference predictive
vector remains within
$L^{-2}+C(L+1)/k\le CL^{-2}$
of $\theta_h$ along every continuation record.
Construct the true and auxiliary records jointly. While they agree,
use the same policy randomization, chance moves, and draws from
sources that are not heavily sampled. At a heavily sampled source,
let $p$ and $q$ be the true and reference predictive distributions.
Draw the outcomes jointly with marginals $p$ and $q$, assigning
probability $\min\{p_a,q_a\}$ to both draws equaling $a$, for each $a$.
$1-\sum_a\min\{p_a,q_a\}=\TV(p,q)\le CL^{-2}$.
The stopping rule makes the same decision while the records agree.
Summing over at most $M\lfloor L\rfloor$ source visits bounds the
probability that the records differ, and hence their total variation
distance, by $C/L$.

Let $Q_y^D$ be the reference predictive record law. In the ratio $\mathcal A_y(w)/Q_y^D(w)$, the policy, known randomization, and heavily sampled source factors cancel. What remains is a product, over the other sources, of the ratios in Lemma~\ref{lem:ratio} with $p=\theta_h$, $n=n_h(y)$, and $m=m_h(w)$, the vector of counts that $w$ adds for source $h$. Since $w$ is short, the lemma bounds each log factor by $N_h(y)\KL(\widehat\theta_h(y)\Vert\theta_h)+C$. Since $y$ is $L$-good,
their sum is at most $Mx_L+C=(\log L)/6+C$. Also
$Q_y\ge K^{-1}Q_y^D$ by \eqref{eq:posterior}. Thus
\begin{equation}\label{eq:auxiliarycomparison}
 \TV\bigl(\mathbb P_\theta(\cdot\mid y),\mathcal A_y\bigr)\le C/L,
 \qquad \mathcal A_y(w)\le CL^{1/6}Q_y(w).
\end{equation}
Records with zero auxiliary probability satisfy the second
inequality automatically.

Apply the first inequality in \eqref{eq:auxiliarycomparison}
to $\Xi(w)\1\{q_{yw}(B)\ge\kappa\}$, which lies in $[0,L]$.
Since this function takes values in $[0,L]$, changing the record
law changes its expectation by at most $L(C/L)=C$. 
On the indicated event, Bayes' rule gives
$Q_y(w)\le\kappa^{-1}Q_y^B(w)$. The second inequality in
\eqref{eq:auxiliarycomparison} now yields
\eqref{eq:epochcomparison}. Here $q_{yw}$ is always the
\end{proof}

\subsection{Combining the Bounds}
\begin{proof}[Proof of Theorem~\ref{thm:learning}]
Fix a payoff assignment $g\in\mathcal G_U$. Payoffs, losses, and optimal
policies below are those under $g$, which we omit from the notation.
Fix a true parameter $\theta$, an optimal policy $\sigma$,
and $\bar\ell,\zeta>0$. Call a period a mistake if
$\ell(s_t,\theta)\ge\bar\ell$. By Lemma~\ref{lem:lipschitz},
with $\eta=\bar\ell/(4MU)$ and $B=B_\eta(\theta)$, a
mistake has loss at least $\bar\ell/2$ at every parameter
in $B$. Take $\kappa>0$ from Lemma~\ref{lem:mass}.

\paragraph{Step 1: Epochs.}
Partition the periods before an integer cutoff $T$ into epochs.
At the start of each epoch record the count $n_{ha}$ of each
cell, and end the epoch when some cell's increment reaches
$n_{ha}+1$, after $\lfloor H\rfloor$ periods, or at $T$,
whichever comes first. Each source is used at most once per
period, so every epoch record is short relative to its initial
history. The number of epochs is at most
\begin{equation}\label{eq:epochs}
 J_T=2+T/\lfloor H\rfloor
       +d_{\mathrm{tot}}\lfloor\log_2(T+1)\rfloor,
 \qquad d_{\mathrm{tot}}=\sum_h|A_h|.
\end{equation}
At most $T/\lfloor H\rfloor$ epochs reach the length cap,
and at most one ends at $T$. Every other epoch doubles some
cell's count plus one. Each cell can do this at most
$\lfloor\log_2(T+1)\rfloor$ times.

\paragraph{Step 2: Mistakes within an epoch.}
For an epoch starting at $y$, let $\Bad_{\bar\ell}(w)$ count
its mistakes. The epoch lasts at most $\lfloor H\rfloor$
periods and $\beta^{\lfloor H\rfloor}\ge1/4$. Therefore,
at every $\vartheta\in B$, its discounted loss is at least
$(\bar\ell/8)\Bad_{\bar\ell}(w)$. Integrating over $B$,
using nonnegativity of regret and \eqref{eq:bayesregret}, gives
\begin{equation}\label{eq:bayesmistakes}
 \int_B\mathbb E_\vartheta[
 \Bad_{\bar\ell}(w)\mid y]q_y(d\vartheta)
 \le C_{\bar\ell}H^{2/3}.
\end{equation}
At an $H$-good initial history, apply
\eqref{eq:epochcomparison} with $L=H$ and
$\Xi=\Bad_{\bar\ell}$. Equation~\eqref{eq:bayesmistakes}
then implies
\begin{equation}\label{eq:epochmistakes}
 \mathbb E_\theta\!\left[
 \Bad_{\bar\ell}(w)\1\{q_{yw}(B)\ge\kappa\}\mid y\right]
 \le C_{\bar\ell,\zeta}H^{5/6}.
\end{equation}

\paragraph{Step 3: Summing over epochs.}
Let $\Omega_0$ be the event that every history is $H$-good
and every posterior assigns $B$ mass at least $\kappa$.
Lemmas~\ref{lem:mass} and~\ref{lem:goodhistories} give
$\mathbb P_\theta(\Omega_0^c)\le\zeta+p_H$, independently
of $T$. Whether an epoch starts at an $H$-good history is measurable
at its start. On $\Omega_0$, every epoch starts at such a
history and ends with $q_{yw}(B)\ge\kappa$. Summing \eqref{eq:epochmistakes} by conditional
expectation over the at most $\lceil J_T\rceil$ epochs gives
\begin{equation}\label{eq:mistakes}
 \mathbb E_\theta\!\left[
 \1_{\Omega_0}\sum_{t<T}\1\{\ell(s_t,\theta)\ge\bar\ell\}
 \right]
 \le C_{\bar\ell,\zeta}J_TH^{5/6}.
\end{equation}

\paragraph{Step 4: Discounted frequency of mistakes.}
Take $\zeta<1$, and set  $T=\lceil\log(1/\zeta)/\psi\rceil$, where
$\psi=1-\lambda\le H^{-1}$. Splitting at $T$ and according
to $\Omega_0$, and using \eqref{eq:mistakes}, bounds the
discounted mistake frequency by
$\zeta+p_H+C_{\bar\ell,\zeta}\psi J_TH^{5/6}+\lambda^T$.
Here $\lambda^T\le e^{-\psi T}\le\zeta$ and $\psi T\le\log(1/\zeta)+1$.
Moreover,
$\psi(1+\log(T+1))\le C_\zeta H^{-1}(1+\log H)$ uniformly
over $\psi\le H^{-1}$: use $T+1\le(\log(1/\zeta)+2)/\psi$ and
monotonicity of $x\log(1/x)$ for sufficiently small $x$.
Equation~\eqref{eq:epochs} therefore gives
\begin{equation}\label{eq:frequency}
 \psi\sum_{t\ge0}\lambda^t
 \mathbb P_\theta\{\ell(s_t,\theta)\ge\bar\ell\}
 \le2\zeta+p_H
       +C_{\bar\ell,\zeta}H^{-1/6}(1+\log H).
\end{equation}
This holds for every $\lambda\ge\beta$.
The constant in \eqref{eq:bayesregret} depends on payoffs
only through $U$. Lemma~\ref{lem:lipschitz} gives the common
neighborhood radius $\eta=\bar\ell/(4MU)$. All posterior,
likelihood, and coupling bounds are independent of payoffs.
Thus \eqref{eq:frequency} holds uniformly over
$g\in\mathcal G_U$ and policies optimal for that array.
Regret is at most
$\bar\ell$ plus $2U$ times the mistake frequency.
By \eqref{eq:goodhistories}, $p_H\to0$. Taking the uniform
limit superior in \eqref{eq:frequency}, then letting
$\zeta\downarrow0$ and $\bar\ell\downarrow0$, establishes
the claimed limit for $\lambda\ge\beta$.

For $0\le\lambda<\beta$ with
$1-\lambda\le C_0(1-\beta)$, nonnegativity of losses
and $\lambda^t\le\beta^t$ give
\begin{equation}\label{eq:shorter-horizon}
 R^g_\lambda(\theta,\sigma)
 \le \frac{1-\lambda}{1-\beta}R^g_\beta(\theta,\sigma)
 \le C_0 R^g_\beta(\theta,\sigma).
\end{equation}
The case $\lambda=\beta$ already proved makes the right side
vanish uniformly over $g$, $\theta$, and policies optimal
under $g$. Together with the result for $\lambda\ge\beta$,
this proves \eqref{eq:uniform}.
\end{proof}

\section[Learning with Deterministic Lifetimes]{Learning with Deterministic Lifetimes}
\label{sec:deterministic}

Maintain the game, feedback, matching, and prior assumptions of Section~\ref{sec:steady}, but suppose each agent lives for  $T$ periods, as in \citet{Fudenberg1993}. A mass $1/T$ of newborn agents enters each role each period, so every age $t\in\{0,\ldots,T-1\}$ has mass $1/T$. Agents know their remaining lifetimes and discount by $\delta\in(0,1]$. Fix each role's prior $q_{i,0}$ independently of $T$ and $\delta$. An age-$t$ agent solves a Bayesian decision problem with $T-t$ periods remaining.

Let $S_i$ be role $i$'s set of pure plans, let
$\sigma_{i,t}(s_i\mid y)$ be its possibly randomized policy at
age $t$ and personal history $y$, and let $\mu_{i,t}$ be the
distribution of those histories. A steady state here consists of
optimal policies and history distributions such that newborn
histories are empty, $\mu_{i,t+1}$ is generated from $\mu_{i,t}$
by the policies and matching, and the aggregate distribution
of pure plans is
\begin{equation}\label{eq:det-aggregate}
 \pi_i(s_i)=\frac1T\sum_{t=0}^{T-1}\sum_y
 \mu_{i,t}(y)\sigma_{i,t}(s_i\mid y).
\end{equation}
Each agent in role $i$ faces independent draws from the other
roles' aggregate distributions $\pi_{-i}$.\footnote{Brouwer's fixed-point theorem implies that steady states exist.} \begin{proposition}[Deterministic lifetimes]\label{prop:det-nash}
Fix the game and priors satisfying Assumption~\ref{asst:1}.
For any sequence $T_n\to\infty$ and $\delta_n\in(0,1]$ with $\delta_n\to1$, every
accumulation point of steady-state profiles $\pi^n$ is a Nash
equilibrium. No restriction on the relative rates of these
limits is required.
\end{proposition}

The proof uses a finite-horizon version of the learning bound.
In the common learning problem, let $\Opt_{\delta,T}(q_0)$ be
all possibly randomized policies that maximize the prior expectation of
$\sum_{t=0}^{T-1}\delta^t u(s_t,\theta)$, and define their
undiscounted average regret by
\begin{equation}\label{eq:det-regret}
 \overline R_{T}(\theta,\sigma)
 =\frac1T\mathbb E_\theta^\sigma
 \left[\sum_{t=0}^{T-1}\ell(s_t,\theta)\right].
\end{equation}

\begin{lemma}\label{lem:det-learning}
For every fixed learning problem and prior satisfying
Assumption~\ref{asst:1},
\begin{equation}\label{eq:det-uniform}
 \lim_{\substack{T\to\infty\\\delta\to1,\ \delta\in(0,1]}}
 \sup_\theta\sup_{\sigma\in\Opt_{\delta,T}(q_0)}
 \overline R_{T}(\theta,\sigma)=0.
\end{equation}
The supremum includes boundary parameters. The same limit holds
uniformly over payoff assignments $g\in\mathcal G_U$ and the policies optimal for each.
\end{lemma}

\begin{proof}
Write $H=(1-\delta)^{-1}$ and $L=\min\{T,H\}$, with
$H=\infty$ at $\delta=1$. It suffices to consider $L\ge2$.
With $r$ periods remaining, the exploration policy used in
Lemma~\ref{lem:bayesregret}, truncated at $r$, gives
\begin{equation}\label{eq:det-bayes}
 \int\mathbb E_\vartheta^\sigma\!\left[
 \sum_{j=0}^{r-1}\delta^j\ell(s_j,\vartheta)\right]q(d\vartheta)
 \le C\min\{r,H\}^{2/3}
\end{equation}
for every posterior $q$ and optimal continuation $\sigma$.
Indeed, sampling each action $m$ times gives a regret bound
$C(m+\min\{r,H\}/\sqrt m)$; take
$m=\lceil\min\{r,H\}^{2/3}\rceil$. If death occurs during
exploration, the entire loss is at most $Cm$.

Fix $\bar\ell,\zeta>0$. Set $\eta=\bar\ell/(4MU)$ and
$B=B_\eta(\theta)$, and take $\kappa$ from
Lemma~\ref{lem:mass}. Partition the $T$ periods into the
epochs of Appendix~\ref{sec:proof}, with length cap
$\lfloor L\rfloor$. Their number is at most
$C(T/L+1+\log(T+1))$. Since
$\delta^{\lfloor L\rfloor}\ge1/4$ and
$\min\{r,H\}\le L$, \eqref{eq:det-bayes} bounds the
Bayesian expected number of mistakes in an epoch, integrated
over $B$, by $C_{\bar\ell}L^{2/3}$. At every $L$-good
initial history, Lemma~\ref{lem:epochcomparison}, applied 
to the number of mistakes in the epoch, therefore
gives the bound $C_{\bar\ell,\zeta}L^{5/6}$ for the true
expected mistake count restricted to records whose terminal
posterior assigns $B$ mass at least $\kappa$.

Extend the policy arbitrarily after death. By
Lemmas~\ref{lem:mass} and~\ref{lem:goodhistories},  the
probability that any initial-goodness or posterior-mass
condition fails is at most $\zeta+p_L$. Summing the epoch
bounds as in \eqref{eq:mistakes} and dividing by $T$ gives
\begin{equation}\label{eq:det-frequency}
 \begin{aligned}
 \frac1T\sum_{t<T}\mathbb P_\theta^\sigma
 \{\ell(s_t,\theta)\ge\bar\ell\}
 &\le\zeta+p_L+
 C_{\bar\ell,\zeta}
 \left(\frac1L+\frac{1+\log(T+1)}T\right)L^{5/6}\\
 &\le\zeta+p_L+
 C_{\bar\ell,\zeta}(1+\log L)L^{-1/6}.
 \end{aligned}
\end{equation}
The second inequality uses $T\ge L$ and the fact that
$(1+\log(T+1))/T$ decreases with $T$.
As $T\to\infty$ and $\delta\to1$, $L\to\infty$.
Regret is at most $\bar\ell$ plus $2U$ times the frequency
in \eqref{eq:det-frequency}. Taking the uniform limit
superior, then letting $\zeta\downarrow0$ and
$\bar\ell\downarrow0$, proves \eqref{eq:det-uniform}.
As in the proof of Theorem~\ref{thm:learning}, the constants depend
on payoffs only through $U$, so the bound is uniform over $g\in\mathcal G_U$.
\end{proof}

Because Lemma~\ref{lem:det-learning} is uniform over payoff assignments, the
argument for Proposition~\ref{prop:design} also covers a buyer who lives
for $T$ periods and a seller who weights these periods equally: the
conclusions in \eqref{eq:D3} hold as $T\to\infty$ and $\delta\to1$ at any
relative rates, including $\delta=1$.

\begin{proof}[Proof of Proposition~\ref{prop:det-nash}]
Fix a steady state and a role $i$. By perfect recall and Kuhn's theorem,
as in the proof of Theorem~\ref{thm:nash}, each
opponent's aggregate mixture of pure plans is realization
equivalent to a behavioral strategy. Independent matching
therefore gives a fixed parameter $\theta_i$ for role $i$'s
learning problem. A pure plan is an action, and opponent
information sets are sources, each visited at most once per
path. The role's policy is optimal for its $T$-period problem
with discount factor $\delta$.

By \eqref{eq:det-aggregate}, its Nash regret is $
 \max_{s_i}u_i(s_i,\pi_{-i})-u_i(\pi_i,\pi_{-i})
 =\overline R_{T}(\theta_i,\sigma_i).$ 
Lemma~\ref{lem:det-learning} makes each role's regret vanish uniformly along the joint limit. Continuity of payoffs gives the conclusion.
\end{proof}

\section{Proofs for the  Applications}

\subsection{Optimal hidden design}\label{app:design}

\begin{proof}[Proof of Proposition~\ref{prop:design}]By Theorem~\ref{thm:learning},  
\begin{equation}\label{eq:D1}
 \varepsilon_\beta:=
 \sup_{\substack{\lambda\in[0,1)\\
                 1-\lambda\le C_0(1-\beta)}}
 \sup_{\theta\in\Theta,\,p\in[0,1]}
 \ \sup_{\sigma\in\Opt_\beta(q_0;g_p)}
 R_\lambda(\theta,p;\sigma)\longrightarrow0.
\end{equation}
Let $\Pi^*_\beta=\sup_{\theta,p}
\Pi_{\lambda_\beta}(\theta,p;\sigma_\beta(p))$.
Write $\bar v=\bar v(\theta)$,
$\phi=\phi_{\lambda_\beta}(\theta,p;\sigma_\beta(p))$,
and $\varepsilon=\varepsilon_\beta$.
By \eqref{eq:D1}, $[p-\bar v]_+\phi\le\varepsilon$, so
\begin{equation}\label{eq:D4}
 \Pi=\mathcal S(\theta)\phi+(p-\bar v)\phi
 \le [\mathcal S(\theta)]_+\phi+\varepsilon
 \le \mathcal S^*\phi+\varepsilon
 \le \mathcal S^*+\varepsilon.
\end{equation}

For a lower bound, choose $\theta^*\in\argmax_\Theta\mathcal S$.
Since $c\ge0$, $\bar v(\theta^*)\ge\mathcal S^*$.
For $\Delta_p\in(0,\mathcal S^*)$, the price
$p=\bar v(\theta^*)-\Delta_p$ lies in $[0,1]$.
At this design and price, \eqref{eq:D1} gives
$\Delta_p(1-\phi)\le\varepsilon$, and hence
$\Pi\ge(\mathcal S^*-\Delta_p)(1-\varepsilon/\Delta_p)$.
If $0<\varepsilon<(\mathcal S^*)^2$, take $\Delta_p=\sqrt{\varepsilon}$.
Since $\mathcal S^*\le1$, this gives
$\Pi^*_\beta\ge\mathcal S^*-2\sqrt{\varepsilon}$.
If $\varepsilon=0$, let $\Delta_p\downarrow0$.
Thus, whenever $\varepsilon_\beta<(\mathcal S^*)^2$,
\begin{equation}\label{eq:D2}
 \mathcal S^*-2\sqrt{\varepsilon_\beta}
 \le \Pi^*_\beta\le\mathcal S^*+\varepsilon_\beta.
\end{equation}

Now consider seller choices $(\theta_\beta,p_\beta)$
whose profits are within $\eta_\beta\to0$ of $\Pi^*_\beta$.
Since $\varepsilon_\beta\to0$, \eqref{eq:D2} implies
$\Pi\to\mathcal S^*$.
Equation~\eqref{eq:D4} gives
$[\mathcal S(\theta_\beta)]_+\ge\Pi-\varepsilon\to\mathcal S^*$.
As $\mathcal S(\theta_\beta)\le\mathcal S^*$ and $\mathcal S^*>0$,
$\mathcal S(\theta_\beta)\to\mathcal S^*$.
Compactness of $\Theta$ and continuity of $\mathcal S$ then imply
$\operatorname{dist}(\theta_\beta,\argmax_\Theta\mathcal S)\to0$.

Also, \eqref{eq:D4} gives
$\mathcal S^*\phi\ge\Pi-\varepsilon\to\mathcal S^*$, so $\phi\to1$.
For prices above mean value,
$[p_\beta-\bar v(\theta_\beta)]_+\phi\le\varepsilon$.
For prices below mean value,
$(\bar v(\theta_\beta)-p_\beta)\phi
=\mathcal S(\theta_\beta)\phi-\Pi\le\mathcal S^*-\Pi$.
Since $\phi\to1$, both price gaps tend to zero.
Thus $p_\beta-\bar v(\theta_\beta)\to0$, completing
\eqref{eq:D3}.
\end{proof}
\subsection{Congestion}\label{app:congestion}

\begin{proof}[Proof of Proposition~\ref{prop:congestion-existence}]
The selected policies depend on the priors, tolls, and $\beta$,
not on the true flow. For each age cutoff $N$, the truncated
right side of \eqref{eq:congestion-fixed-point} is a finite sum
over histories whose probabilities are polynomial in
$\theta(F)$, which is continuous in $F$. The omitted tail in
each coordinate is at most $\gamma^N$ uniformly, so the full
expression is continuous. It maps $\mathcal F$ into itself:
the induced route flows in class $k$ sum to
$\rho_k(1-\gamma)\sum_{t\ge0}\gamma^t=\rho_k$.
Since $\mathcal F$ is nonempty, compact, and convex,
Brouwer's theorem gives a fixed point.
\end{proof}

\begin{proof}[Proof of Proposition~\ref{prop:wardrop}]
By Theorem~\ref{thm:learning} and finiteness of the class
set, there is $\varepsilon(\beta)\to0$ such that
$R_{\lambda}(\theta,\sigma)\le\varepsilon(\beta)$
for every class $k$, $\lambda\in[\beta,1)$, parameter $\theta$,
and $\sigma\in\Opt_\beta(q_{k,0};g^\chi)$. Fix a stationary learning equilibrium $F$ with class policies
$\sigma_k$. At $\theta(F)$, a route's payoff loss is its
expected delay plus toll minus the lowest such cost in its
class. By \eqref{eq:congestion-fixed-point}, $F_r/\rho_k$
is the expected $\gamma$-discounted frequency of route $r$.
Thus class $k$ contributes
$\rho_kR_{\gamma}(\theta(F),\sigma_k)$ to $G_\chi(F)$.
Since $\gamma\ge\beta$ and $\sum_k\rho_k=1$, summing over
classes gives $0\le G_\chi(F)\le\varepsilon(\beta)$.

Continuity of $G_\chi$ implies that every accumulation point
as $\beta\uparrow1$ has zero gap and hence is a Wardrop
equilibrium. Compactness of $\mathcal F$ then implies
convergence in distance to the Wardrop equilibrium set.
\end{proof}

\begin{proof}[Proof of Proposition~\ref{prop:welfare}]
Inequality~\eqref{eq:poa} is the final display in the proof of
\citet[Theorem~6.2]{BlumEvenDarLigett2010}, applied to a constant flow,
whose regret is $G(F)$, with their constant equal to $4/3$ for affine delays.
For the tolls $\chi_e^*=\mathsf a_ex_e^{SO}$, direct expansion
gives
$\mathcal C(F)-\mathcal C^{SO}
=\sum_e[\mathsf t_e(x_e)+\chi_e^*](x_e-x_e^{SO})
\le G_{\chi^*}(F)$.
The inequality follows because every route used by $F^{SO}$
costs at least the cheapest route in its class at $F$.
Optimality of $F^{SO}$ gives the lower bound in
\eqref{eq:tolls}.
Proposition~\ref{prop:wardrop} gives the limiting welfare
claims for stationary learning equilibria.
\end{proof}

\clearpage
\pagenumbering{arabic}                    
\renewcommand{\thepage}{OA-\arabic{page}} 
\onehalfspacing

\section{For Online Publication}

\subsection{Congestion: potential bound and example}\label{sec:OLA_cong}
\paragraph{Potential bound.}
If each $\mathsf t_e$ is nondecreasing, the potential
\[\Phi_\chi(x)=\sum_e\int_0^{x_e}\mathsf t_e(z)\,dz
+\sum_e\chi_ex_e\] is convex
\citep{BeckmannMcGuireWinsten1956}.
Let $x^W$ minimize $\Phi_\chi$ over feasible edge loads,
and let $F^W$ be a route flow realizing $x^W$.
For any feasible route flow $F$ with edge loads $x$,
convexity gives
\begin{equation}\label{eq:beckmann}
 \begin{aligned}
 0\le\Phi_\chi(x)-\Phi_\chi(x^W)
 &\le\nabla\Phi_\chi(x)\cdot(x-x^W)\\
 &=\sum_r(c_r(F)+\chi_r)(F_r-F_r^W)
 \le G_\chi(F).
 \end{aligned}
\end{equation}
The last inequality follows because the routes used by $F^W$
cost at least the cheapest route in their class at $F$.
If every $\mathsf t_e$ is differentiable with
$\mathsf t_e'\ge m>0$, strong convexity and the first-order
condition at $x^W$ give
$\Phi_\chi(x)-\Phi_\chi(x^W)\ge(m/2)\|x-x^W\|_2^2$,
and hence $\|x-x^W\|_2\le\sqrt{2G_\chi(F)/m}$.

\begin{proof}[Claims in Example~\ref{ex:two-link}]
If $\beta<1/3$, any policy starting with $A$ has normalized
Bayesian cost at least $(1-\beta)3/4>1/2$, since subsequent
costs are nonnegative. Always choosing $B$ costs $1/2$ and
leaves beliefs unchanged. Hence every optimal policy chooses
$B$ forever, giving the unique stationary flow $x=0$.

For high patience, consider trying $A$ for three periods,
continuing with $A$ forever if all three delays are zero,
and otherwise switching to $B$. Under the prior, the probability
of three zeros is $\int_0^1 3\theta^2(1-\theta)^3\,d\theta
=1/20$, and the posterior mean after three zeros is $3/7$.
This policy's normalized Bayesian cost is
$3(1-\beta^3)/4+
\beta^3[(1/20)(3/7)+(19/20)(1/2)]
=3/4-71\beta^3/280$,
which is below $1/2$ when $\beta^3>70/71$.
If $B$ were optimal at the initial belief, its lack of
information would imply that the optimal cost $V$ satisfies
$V=(1-\beta)/2+\beta V$, and hence $V=1/2$.
Thus every optimal policy starts with $A$, and the newborn
cohort gives $x\ge1-\gamma>0$ in every stationary equilibrium.

The untolled gap is
$G(x)=(1-x)(1/2-x)$ for $x\le1/2$ and
$G(x)=x(x-1/2)$ for $x\ge1/2$.
Thus $G(x)\ge|x-1/2|/2$, and
Proposition~\ref{prop:wardrop} gives
$|x_{\beta,\gamma}-1/2|\le2\varepsilon(\beta)\to0$.

Total expected travel time satisfies
$\mathcal C(x)=(x-1/4)^2+7/16$, so its unique minimizer
is $x^{SO}=1/4$. With $\chi_A=1/4$ and $\chi_B=0$,
the perceived route costs are $x+1/4$ and $1/2$.
The gap is $(1-x)(1/4-x)$ for $x\le1/4$ and
$x(x-1/4)$ for $x\ge1/4$, so
$G_\chi(x)\ge|x-1/4|/4$.
Proposition~\ref{prop:wardrop} therefore gives
$x_{\beta,\gamma}\to1/4$ and
$\mathcal C(x_{\beta,\gamma})\to7/16$.
Finally, if $\beta<1/2$, any policy starting with $A$
has normalized Bayesian cost at least
$(1-\beta)(3/4+1/4)>1/2$.
Thus every optimal policy chooses $B$ forever.
\end{proof}

\subsection{Example~\ref{ex:bandit-atom}}\label{app:bandit-counterexample}
\begin{proof}
[Proof of the claims in Example~\ref{ex:bandit-atom}]
Let $y^{(n)}$ be the history of $n$ pulls of $A$, all yielding $1/2$, and put $r(\theta)=1-\theta_A(1/2)$. The posterior at $y^{(n)}$ is
\begin{equation}\label{eq:bandit-atom-posterior}
 q_n(d\theta)=\frac{\tfrac9{10}\nu^0(d\theta)
       +\tfrac1{10}(1-r(\theta))^n\nu(d\theta)}{Z_n},
 \qquad Z_n=\tfrac9{10}+\frac1{5(n+1)(n+2)}.
\end{equation}
Indeed, under $\nu$, $r$ has density $2r$ on $[0,1]$, so
\begin{equation}\label{eq:bandit-atom-integrals}
 \int(1-r)^n\,d\nu=\frac2{(n+1)(n+2)},\qquad
 \int\frac{(1-r)^n}{r}\,d\nu=\frac2{n+1}.
\end{equation}
The second integral ignores the $\nu$-null set $\{r=0\}$.

Fix $n$ and a continuation policy $\sigma$ that chooses $B$ first. Construct $\widetilde\sigma$ by choosing $A$ until its first reward different from $1/2$, then running $\sigma$ from its original internal history $y^{(n)}$, ignoring the waiting-phase observations. If this reward never occurs, choose $A$ forever. Write $v_\beta(\sigma;\theta)$ for the normalized continuation payoff and $b=1-\beta$.

At $\theta^0$, the comparison policy earns $1/2$ forever, whereas $\sigma$ earns zero initially and at most $1/2$ thereafter. Hence its gain is at least $b/2$. For $r(\theta)>0$, let $\tau$ count the $A$ pulls through the first reward different from $1/2$. Conditional on $\theta$, $\tau$ is geometric with mean $1/r(\theta)$, and the draws used after $\tau$ are fresh. Waiting-phase rewards are nonnegative and $0\le v_\beta(\sigma;\theta)\le1$, so
\begin{equation}\label{eq:bandit-atom-delay}
 \begin{aligned}
 v_\beta(\widetilde\sigma;\theta)-v_\beta(\sigma;\theta)
 &\ge-\bigl(1-\mathbb E_\theta[\beta^\tau]\bigr)v_\beta(\sigma;\theta)\\
 &\ge-b\mathbb E_\theta[\tau]=-\frac b{r(\theta)}.
 \end{aligned}
\end{equation}
Integrating the two bounds against \eqref{eq:bandit-atom-posterior} gives
\begin{equation}\label{eq:bandit-atom-gain}
 \int\bigl[v_\beta(\widetilde\sigma;\theta)-v_\beta(\sigma;\theta)\bigr]q_n(d\theta)
 \ge\frac b{Z_n}\left[\frac9{20}-\frac1{5(n+1)}\right]
 \ge\frac b{4Z_n}>0.
\end{equation}
Thus every optimal continuation at $y^{(n)}$ chooses $A$ with probability one; any positive probability of $B$ can be strictly improved by replacing that branch. Starting at $y^{(0)}$, these histories have positive prior-predictive probability because of the atom. Optimality therefore requires optimal continuations there. At $\theta^*$, choosing $A$ always yields $1/2$, so induction shows that every optimal policy chooses $A$ forever. More generally, at any $\theta$ with $r(\theta)>0$, every optimal policy chooses $A$ until its first reward different from $1/2$, which arrives at a geometric time $\tau$ with mean $1/r(\theta)$. Its normalized regret at $\theta$ is therefore at least $\ell(A,\theta)\bigl(1-\mathbb E_\theta[\beta^\tau]\bigr)=\ell(A,\theta)\,b/(b+\beta r(\theta))$, which is at least $\ell(A,\theta)/2$ whenever $r(\theta)\le b$. Since $\ell(A,\theta)\to1/2$ as $\theta\to\theta^*$, regret does not vanish uniformly on any neighborhood of $\theta^*$, even with $\theta^*$ excluded.

Finally, every feasible finite action--reward history has a likelihood strictly positive in the relative interior of $\Theta$. Its posterior therefore retains a diffuse component assigning positive probability to every nonempty relatively open set. Full support holds after every finite history.

\end{proof}

\subsection{Example~\ref{ex:1}}\label{app:counterexample}
\begin{proof}[Proof of the steady-state claim]
Fix a history of $n$ observations of (Safe, Left, Stop). Player~1's likelihood is $(1-p_{\mathrm P})^n(1-\xi)^n$, and Player~2's is $(1-p_{\mathrm R})^n(1-\xi)^n$. Consequently,
\begin{equation}\label{eq:counter-posterior}
 \begin{gathered}
 p_{\mathrm P}\sim\operatorname{Beta}(2,n+1)\quad\text{for Player~1},\qquad
 p_{\mathrm R}\sim\operatorname{Beta}(2,n+1)\quad\text{for Player~2},\\
 \xi\sim\operatorname{Beta}(1,n+1)\quad\text{for either player}.
 \end{gathered}
\end{equation}
In each posterior, $p_{\mathrm P}$ or $p_{\mathrm R}$ remains independent of Player~3's strategy, and the conditional density of $(p_0,p_1)$ given $\xi$ is unchanged. Write $\mathbb E_n$ for expectation under the relevant player's posterior.

Let $Z_1=P$ and $Z_2=1-P$. Under the corresponding player's beliefs, the conditional laws of $Z_i$ given $\xi$ agree, and
\begin{equation}\label{eq:counter-moments}
 A(\xi):=\mathbb E_n[(\bar u-Z_i)_+\mid\xi]\ge\frac{1-\xi/2}{3},\qquad
 B(\xi):=\mathbb E_n[(Z_i-\bar u)_+\mid\xi]\le\frac{9\xi}{100}.
\end{equation}
In the component with density $f_{\mathrm{low}}\otimes f_{\mathrm{low}}$ (reflected for Player~2), $Z_i\le1/5$ and its mean is $1/15$, so the expected loss relative to $\bar u=2/5$ is $1/3$ and the positive gain is zero. In the uniform component, convexity gives $\mathbb E[((1-\xi)U_0+\xi U_1-\bar u)_+]\le\mathbb E[(U_0-\bar u)_+]=9/50$ for uniform $U_0,U_1$. Multiplying by its weight $\xi/2$ proves the bounds.

\paragraph{Player~1.}
Fix $\beta=\delta\gamma<1$ and put $b=1-\beta$. Consider a policy $\sigma$ that chooses Risky first. A comparison policy $\widetilde\sigma$ chooses Safe until the first Pass, uses that match's pair $(x,y)$ as $\sigma$'s initial Risky observation, and then follows its continuation. It ignores all preceding observations and uses $\sigma$'s own internal history and clock. Conditional on fixed parameters, the revealing pair has the same distribution as under Risky, independently of the waiting time $\tau_1\sim\operatorname{Geom}(p_{\mathrm P})$ on $\{1,2,\ldots\}$. If $p_{\mathrm P}=0$, the comparison chooses Safe forever. For $p_{\mathrm P}>0$, $
 z_1(p_{\mathrm P})=\mathbb E_{p_{\mathrm P}}[\beta^{\tau_1}]=\frac{\beta p_{\mathrm P}}{b+\beta p_{\mathrm P}}\le\beta.$ Let $W$ be the normalized value of $\sigma$'s continuation, averaged over its initial observation, at the fixed parameters. The comparison's continuation has the same value, because the virtual initial observation it is fed has the same distribution at the fixed parameters; $W$ is not a posterior value at the comparison's actual history. Since $W\le\max\{\bar u,P\}$,
\begin{equation}\label{eq:counter-gain}
 \begin{aligned}
 v_{\widetilde\sigma}-v_\sigma
 &=\bar u(1-z_1)+z_1W-bP-\beta W\\
 &\ge b\left[(\bar u-P)_+-\frac{(P-\bar u)_+}{p_{\mathrm P}}\right].
 \end{aligned}
\end{equation}
For $P\le\bar u$, substitute $W\le\bar u$. For $P>\bar u$, substitute $W\le P$ and use $1-z_1=b/(b+\beta p_{\mathrm P})\le b/p_{\mathrm P}$. The posterior assigns probability zero to $p_{\mathrm P}=0$. Its independence and the identities $\mathbb E_n[\xi]=1/(n+2)$ and $\mathbb E_n[1/p_{\mathrm P}]=n+2$ yield
\begin{equation}\label{eq:counter-strict}
 \mathbb E_n[v_{\widetilde\sigma}-v_\sigma]
 \ge b\left[\frac13-\frac1{6(n+2)}-\frac9{100}\right]
 =b\frac{73n+96}{300(n+2)}>0.
\end{equation}
Thus every optimal policy chooses Safe at this history.

\paragraph{Player~2.}
Consider a policy $\sigma$ whose initial plan is Pass. The comparison initially selects Stop. If Player~1 chooses Risky in this first match, Player~2's plan affects neither payoffs nor observations. The comparison feeds the actual terminal observation to $\sigma$, together with its original virtual Pass plan, and follows the same continuation. This branch has zero payoff difference.

If Player~1 chooses Safe, the comparison receives $\bar u$, ignores the diagnostic observation $x$, and continues choosing Stop until the first future Risky match. It uses that match's $(x,y)$ as $\sigma$'s original Safe--Pass observation and follows its continuation, ignoring all intervening observations.

Condition on fixed parameters and first-period Safe, without conditioning on the ignored diagnostic observation. The eventual Risky pair has the same law as the original Safe--Pass pair, independently of the delay. For $p_{\mathrm R}>0$, $ \tau_2=1+\operatorname{Geom}(p_{\mathrm R})$, $  z_2(p_{\mathrm R})=\mathbb E_{p_{\mathrm R}}[\beta^{\tau_2}]=\dfrac{\beta^2p_{\mathrm R}}{b+\beta p_{\mathrm R}}$, and $\beta-z_2(p_{\mathrm R})=\dfrac{\beta b}{b+\beta p_{\mathrm R}}.$ At $p_{\mathrm R}=0$, the comparison keeps choosing Stop. A Stop plan has mean payoff $\bar u$, and a Pass plan has mean $p_{\mathrm R}\bar u+(1-p_{\mathrm R})(1-P)$, so the normalized continuation value $W$, defined at the fixed parameters as for Player~1, satisfies $W\le\bar u+(1-p_{\mathrm R})(1-P-\bar u)_+$. On the first-period Safe branch, therefore,
\begin{equation}\label{eq:counter-gain2}
 \begin{aligned}
 v_{\widetilde\sigma}-v_\sigma
 &=\bar u(1-z_2)+z_2W-b(1-P)-\beta W\\
 &\ge b\left[(\bar u-1+P)_+-\frac{(1-P-\bar u)_+}{p_{\mathrm R}}\right].
 \end{aligned}
\end{equation}
For $1-P\le\bar u$, use $W\le\bar u$. Otherwise the coefficient on the loss is $b+(\beta-z_2)(1-p_{\mathrm R})=b/(b+\beta p_{\mathrm R})\le b/p_{\mathrm R}$. Multiplying by the first-period Safe probability $1-p_{\mathrm R}$ and averaging gives
\begin{equation}\label{eq:counter-strict2}
 \begin{aligned}
 \mathbb E_n[v_{\widetilde\sigma}-v_\sigma]
 &\ge b\mathbb E_n\left[(1-p_{\mathrm R})\left(A(\xi)-\frac{B(\xi)}{p_{\mathrm R}}\right)\right]\\
 &\ge b\left[\frac{n+1}{n+3}\frac{2n+3}{6(n+2)}
             -\frac{9(n+1)}{100(n+2)}\right]\\
 &=b\frac{(n+1)(73n+69)}{300(n+2)(n+3)}>0.
 \end{aligned}
\end{equation}
Here $\mathbb E_n[1-p_{\mathrm R}]=(n+1)/(n+3)$ and $\mathbb E_n[(1-p_{\mathrm R})/p_{\mathrm R}]=n+1$. Thus every optimal policy chooses Stop. For either player, a randomized policy assigning positive probability to the inferior initial plan is strictly improved by replacing only that branch.

\paragraph{Player~3 and stationarity.}
Player~3's actions affect neither her payoff nor her information about $p_{\mathrm R},p_{\mathrm P}$: Player~1 moves before $x$, Player~2 does not observe $x$, and $y$ affects neither the route nor Player~3's payoff. Every policy for Player~3 is therefore optimal. Prescribe $x=0$ and a fair choice of $y$ at each final information set. Complete the other players' policies optimally at histories outside the proposed state. Every match then ends at (Safe, Left, Stop). The geometric age distribution supplies consistent stationary history distributions for every $(\delta,\gamma)$; each of Players~1 and~2 has true stage regret $1/2-\bar u=1/10$.

\paragraph{Deterministic lifetimes.}
Let $r\ge1$ be the remaining lifetime and let $\delta\in(0,1]$. Use the same comparisons, truncated at death, with unnormalized payoffs. When a revealing match occurs, the comparison runs $\sigma$'s continuation on its original internal clock for the available periods.

For Player~1, put $d=P-\bar u$. Conditional on fixed parameters, each expected continuation payoff above $\bar u$ lies between $\min\{0,d\}$ and $\max\{0,d\}$. Delaying and truncating these terms cannot hurt when $d\le0$, so the improvement is at least $-d$. If $d>0$, the loss is at most
\begin{equation}\label{eq:counter-finite1}
 d\,\mathbb E_{p_{\mathrm P}}\!\left[\sum_{t=0}^{\min\{\tau_1,r\}-1}\delta^t\right]
 \le d\,\mathbb E_{p_{\mathrm P}}[\tau_1]=\frac d{p_{\mathrm P}},\qquad \tau_1\sim\operatorname{Geom}(p_{\mathrm P}).
\end{equation}
Thus \eqref{eq:counter-gain} holds without the factor $b$.

For Player~2, the first-period Risky branch again has zero difference. Conditional on first-period Safe, put $d=1-P-\bar u$. Each expected continuation payoff above $\bar u$ lies between $\min\{0,(1-p_{\mathrm R})d\}$ and $\max\{0,(1-p_{\mathrm R})d\}$. For $d\le0$ the improvement is at least $-d$. For $d>0$, let $\tau_2=1+\operatorname{Geom}(p_{\mathrm R})$. The loss is at most
\begin{equation}\label{eq:counter-finite2}
 d+(1-p_{\mathrm R})d\,\mathbb E_{p_{\mathrm R}}\!\left[\sum_{t=1}^{\min\{\tau_2,r\}-1}\delta^t\right]
 \le d+(1-p_{\mathrm R})d\,\mathbb E_{p_{\mathrm R}}[\tau_2-1]=\frac d{p_{\mathrm R}}.
\end{equation}
The gains in \eqref{eq:counter-strict} and \eqref{eq:counter-strict2} remain positive without $b$ at every age. Uniform age weights give the claimed steady state for every lifetime and discount factor.
\end{proof}
\begin{proof}[Prior properties and the calculations in
Figure~\ref{fig:prior-dependence}]
Integrating either density in \eqref{eq:counter-prior}
over $(p_0,p_1)$ gives one, so $\xi$ is uniform.
Integrating over $\xi$ and the other final-choice probability
gives marginal density $3f_{\mathrm{low}}(p_x)/4+1/4$ under $\Lambda_+$ and
$3f_{\mathrm{low}}(1-p_x)/4+1/4$ under $\Lambda_-$.
These densities are continuous and bounded away from zero.
The remaining marginals are beta distributions, so every
one-dimensional marginal satisfies Condition~$\mathcal P$.

After $n$ observations of (Safe, Left, Stop),
\eqref{eq:counter-posterior} gives
$\mathbb E_n[\xi]=1/(n+2)$.
The conditional distribution of $(p_0,p_1)$ given $\xi$
is unchanged, so the posterior weight on its uniform
component is $\mathbb E_n[\xi]/2=1/[2(n+2)]$.
For Player~1, the conditional mean of
$P=(1-\xi)p_0+\xi p_1$ is
$(1-\xi/2)/15+\xi/4=1/15+13\xi/60$.
Hence
$\mathbb E_n[P]=1/15+13/[60(n+2)]$.
Since $f_{\mathrm{low}}$ is supported on $[0,1/5]$,
$\mathbb P_n(p_0>2/5)=(3/5)\mathbb E_n[\xi]/2
=3/[10(n+2)]$.
These are the two curves in
Figure~\ref{fig:prior-dependence}.
Under the product of the same one-dimensional prior
marginals, observations of $x$ leave the distribution of
$p_0$ unchanged, and this probability remains
$(1/4)(3/5)=3/20$.

Equation~\eqref{eq:counter-posterior} also gives
$\mathbb E_n[1/p_{\mathrm P}]=\mathbb E_n[1/p_{\mathrm R}]=n+2$.
Thus, counting from the next match, the posterior expected
number of matches until the other player reveals $(x,y)$
is $n+2$.

To verify that \eqref{eq:prior} fails, fix rectangles
$I_-\subset(0,1/5)^2$ and $I_+\subset(2/5,1)^2$
of positive area.
Under $\Lambda_+$, the probabilities of
$(0,\varepsilon)\times I_-$ and
$(0,\varepsilon)\times I_+$ have orders $\varepsilon$
and $\varepsilon^2$, respectively, as $\varepsilon\downarrow0$.
The ratio of the second probability to the first therefore
tends to zero.
Under any product of beta laws with positive parameters,
the ratio is a positive constant independent of $\varepsilon$,
because the factor for $\xi$ cancels.
A density bounded above and away from zero relative to
such a product would keep the ratio bounded away from
zero. This contradicts \eqref{eq:prior}.
The argument for $\Lambda_-$ follows by reflection.
\end{proof}

\subsection{Finite mixtures of product Dirichlets}\label{sec:mixtures}
The same estimates allow a finite latent mixture of product priors. This includes mixtures with different boundary exponents that need not be comparable to any single product Dirichlet. Mixtures of Dirichlets also appear in \citet[Remark~1]{fudenberg2017rare}.

\begin{proposition}[Finite mixtures]\label{prop:mixtures}
Theorems~\ref{thm:learning} and~\ref{thm:nash},
Proposition~\ref{prop:det-nash}, Lemma~\ref{lem:det-learning},
and their corollaries remain valid if Assumption~\ref{asst:1}
is replaced by
\begin{equation}\label{eq:mixture-prior}
 \begin{gathered}
 q_0(d\theta)=f(\theta)D(d\theta),\qquad
 D=\sum_{j=1}^J \omega_jD^j,\qquad
 D^j=\bigotimes_h\Dir(\alpha_h^j),\\
 \omega_j>0,\quad\sum_j\omega_j=1,\quad\alpha_{ha}^j>0,\qquad
 0<c_-\le f\le c_+<\infty\quad D\text{-a.e.}
 \end{gathered}
\end{equation}
Here $J$ is finite, $f$ is measurable, and the prior is fixed independently of the discount factors and lifetime.
\end{proposition}

\begin{proof}
At every complete or partial history, the reference posterior
is $D_y=\sum_j\omega_j(y)D_y^j$, where
$D_y^j=\bigotimes_h\Dir(\alpha_h^j+n_h(y))$ and
$\omega_j(y)$ are the posterior mixture weights. The
comparison $K^{-1}\le dq_y/dD_y\le K$, with
$K=c_+/c_-$, still holds.

For posterior mass, apply Lemma~\ref{lem:mass} to each
product component with failure probability $\zeta/J$.
On the intersection of these events, every component assigns
$B_\eta(\theta)$ mass at least a common $\kappa_0>0$ in
every period. Their mixture does too, so
$q_y(B_\eta(\theta))\ge K^{-1}\kappa_0$ with probability
at least $1-\zeta$, uniformly over parameters and policies.

The good-history event in Lemma~\ref{lem:goodhistories}
does not depend on the prior. At an $L$-good history,
construct the auxiliary law $\mathcal A_y^j$ of
Lemma~\ref{lem:epochcomparison} separately for each component,
using that component's reference predictive draws for heavily
sampled sources. Let $Q_y^j$ be its full reference predictive
record law. The same proof gives
$\TV(\mathbb P_\theta(\cdot\mid y),\mathcal A_y^j)\le C_j/L$ and
$\mathcal A_y^j(w)\le C_jL^{1/6}Q_y^j(w)$.
The constants can be replaced by their maximum because $J$
is finite.

Mix these auxiliary laws using the weights at the initial
history: $\mathcal A_y=\sum_j\omega_j(y)\mathcal A_y^j$. Convexity of total
variation gives $\TV(\mathbb P_\theta(\cdot\mid y),\mathcal A_y)\le C/L$.
Also $Q_y^D=\sum_j\omega_j(y)Q_y^j$, so
$\mathcal A_y(w)\le CL^{1/6}Q_y^D(w)\le CKL^{1/6}Q_y(w)$.
Thus \eqref{eq:auxiliarycomparison} and the terminal
posterior-mass argument remain valid, proving
\eqref{eq:epochcomparison} for mixture priors.

The Bayesian regret bounds are independent of the prior.
The remaining steps in the proofs of Theorem~\ref{thm:learning}
and Lemma~\ref{lem:det-learning}
therefore apply unchanged, as do their steady-state consequences.
\end{proof}

\end{document}